\documentclass[12pt]{article}
\title{Two-dimensional Brownian vortices}

\usepackage{amsthm,amsmath,amssymb,a4wide,epsf}
\usepackage{color}

cm \hsize=18 true cm \topmargin=-2cm \oddsidemargin=-0.5cm
\def\mb#1{\setbox0=\hbox{$#1$}\kern-.025em\copy0\kern-\wd0
\kern-0.05em\copy0\kern-\wd0\kern-.025em\raise.0233em\box0}

\begin{document}

\author{Pierre-Henri Chavanis}
\maketitle
\begin{center}
Laboratoire de Physique Th\'eorique (CNRS UMR 5152), \\
Universit\'e
Paul Sabatier,\\ 118, route de Narbonne, 31062 Toulouse Cedex 4, France\\
E-mail: {\it chavanis{@}irsamc.ups-tlse.fr\\
 }
%\date{}
\vspace{0.5cm}
\end{center}

\begin{abstract}
We introduce a stochastic model of two-dimensional Brownian vortices
associated with the canonical ensemble. The point vortices evolve
through their usual mutual advection but they experience in addition a
random velocity and a systematic drift generated by the system
as a whole. The statistical equilibrium state of this stochastic model
is the Gibbs canonical distribution. We consider a single species
system and a system made of two types of vortices with positive and
negative circulations. At positive temperatures, like-sign vortices
repel each other (``plasma'' case) and at negative temperatures, like-sign
vortices attract each other (``gravity'' case). We derive the stochastic
equation satisfied by the exact vorticity field and the Fokker-Planck
equation satisfied by the $N$-body distribution function. We present
the BBGKY-like hierarchy of equations satisfied by the reduced
distribution functions and close the hierarchy by considering an
expansion of the solutions in powers of $1/N$, where $N$ is the number
of vortices, in a proper thermodynamic limit. For spatially
inhomogeneous systems, we derive the kinetic equations satisfied by
the smooth vorticity field in a mean field approximation valid for
$N\rightarrow +\infty$. For spatially homogeneous systems, we study
the two-body correlation function, in a Debye-H\"uckel approximation
valid at the order $O(1/N)$. The results of this paper can also apply
to other systems of random walkers with long-range interactions such
as self-gravitating Brownian particles and bacterial populations
experiencing chemotaxis. Furthermore, for positive temperatures, our
study provides a kinetic derivation, from microscopic stochastic
processes, of the Debye-H\"uckel model of electrolytes.

\end{abstract}

\maketitle

\section{Introduction}
\label{sec_introduction}

Systems with long-range interactions are numerous in nature and share
a lot of peculiar properties \cite{dauxois}. For example, it was
recognized long ago that, for systems with unshielded long-range
interactions, the statistical ensembles may be inequivalent at the
thermodynamic limit \cite{lbw,thirring}. Therefore, we must carefully
define the system under consideration before applying a statistical
description. If we consider an {\it isolated} system of particles in
interaction, where the microscopic dynamics is governed by Hamiltonian
equations, the proper statistical ensemble is the microcanonical
ensemble.  Alternatively, if we consider a {\it dissipative} system,
where the microscopic dynamics is governed by stochastic Langevin
equations, the proper statistical ensemble is the canonical
ensemble. A general discussion of the dynamics and thermodynamics of
Hamiltonian and Brownian systems with long range interactions has been
given in \cite{dauxois} and in \cite{hb1,hb2,hb3,hb4,prep} (denoted
Papers I-V).  To any Hamiltonian system with long range interactions,
we can associate a Brownian system with long-range interactions by
adding a friction force and a stochastic force in the equations of
motion.  For example, the self-gravitating Brownian gas studied in
\cite{crs} is the canonical version of a stellar system
\cite{paddy} and the Brownian Mean Field (BMF) model studied in \cite{cvb} is
the canonical version of the Hamiltonian Mean Field (HMF) model
\cite{antoni}. In order to emphasize the similarities and the differences
between Hamiltonian and Brownian systems, it is of great conceptual
interest to study these systems in parallel.

The purpose of the present paper is to extend this study to the case
of point vortices in two-dimensional hydrodynamics. Basically, a gas
of point vortices constitutes a Hamiltonian system described by the
Kirchhoff-Hamilton equations of motion where the coordinates $(x,y)$
of the point vortices are canonically conjugate
\cite{kirchhoff,newton}. A statistical mechanics of this system has
been developed by several authors
\cite{jm,mj,et,kida,pl,lp,caglioti,es,kl} following the pioneering
work of Onsager
\cite{onsager}. For this system, the proper thermodynamic limit is
such that the number of point vortices $N\rightarrow +\infty$ while
their circulation tends to zero as $\gamma\sim 1/N$ and the area of
the system remains finite $V\sim 1$. Assuming ergodicity, the system
should reach, for $t\rightarrow +\infty$, a statistical equilibrium
state described by the mean-field Boltzmann distribution. In order to
vindicate (or not) this result and determine the timescale of the
relaxation, we must develop a kinetic theory of point vortices. A
general kinetic equation describing the collisional relaxation of point
vortices has been derived in \cite{kin} from the Liouville equation by
using a projection operator formalism. In a more recent paper
\cite{bbgky}, we have re-derived this kinetic equation from a BBGKY-like
hierarchy \footnote{A similar approach has been developed
simultaneously in \cite{sano}.} by using an expansion of the solutions
in powers of $1/N$ for $N\rightarrow +\infty$, and we have shown that
this equation is valid at the order $O(1/N)$. In the limit
$N\rightarrow +\infty$, the correlations between point vortices can be
neglected and the kinetic equation reduces to the 2D Euler
equation. This is the counterpart of the Vlasov equation in plasma
physics. On the other hand, by considering an axisymmetric evolution,
we have obtained an explicit expression of the collisional current at
the order $O(1/N)$ taking into account two-body correlations
\cite{kin,bbgky}. This leads to a kinetic equation that is the counterpart of the Landau equation in plasma physics.  Collective effects can be included by developing a
quasilinear theory of the Klimontovich equation \cite{dn,dubin} leading to a kinetic equation analogous to the Lenard-Balescu equation in plasma physics. It
was numerically shown in \cite{cl} that the  kinetic
equation valid at the order $O(1/N)$ does {\it not} in general relax
towards the Boltzmann distribution because of the absence of
resonances. This implies that the collisional relaxation time is
larger than $N t_D$ (where $t_D$ is the dynamical time). However, its
precise scaling with $N$ remains to be determined. This demands to
develop a kinetic theory at the order $1/N^2$ or smaller by taking
into account three-body or higher correlations. This extension has not
yet been done. Therefore, for the moment, there is no theoretical
proof (coming from the kinetic theory) that the Hamiltonian point
vortex gas relaxes towards the Boltzmann distribution for long times.

In this paper, we shall study a {\it different} model of point vortex
dynamics. Our aim is to introduce a stochastic model of point vortices
associated with the canonical ensemble. In other words, we wish to
construct a dynamical model of point vortices that leads to the Gibbs
canonical distribution at statistical equilibrium. These vortices will
be called {\it Brownian vortices} to distinguish them from usual
vortices described by Hamiltonian equations (leading to the
microcanonical distribution at statistical equilibrium). By analogy
with usual Brownian theory, if we want to introduce a canonical
description, the first idea is to modify the usual point vortex model
by adding a stochastic term in the Kirchhoff-Hamilton equations of
motion. This extension has already been considered by Marchioro \&
Pulvirenti \cite{mp} in a different context. Their motivation was to
take into account viscous effects in the point vortex dynamics. They
considered a system of point vortices evolving through their usual
mutual advection and experiencing, in addition, a random velocity
whose strength is proportional to the square root of the viscosity.
They showed that, in a mean-field limit valid for $N\rightarrow
+\infty$, the evolution of the smooth vorticity field is described by
the Navier-Stokes equation (instead of the 2D Euler equation).  This
model introduces a source of dissipation but it does not lead to the
canonical distribution at statistical equilibrium (the Navier-Stokes
equation leads to the trivial state $\omega=0$ for $t\rightarrow
+\infty$). If we want to obtain the canonical distribution, we need to
introduce, in addition, a drift velocity produced by the vortices as a
whole. This drift velocity, which is perpendicular to the advective
velocity, is the counterpart of the dynamical friction in usual
Brownian theory. The drift coefficient, similar to a mobility, is
determined so as to yield the Boltzmann factor at statistical
equilibrium. It is related to the viscosity and to the inverse
temperature by a sort of Einstein relation.

This model of Brownian vortices was introduced in a Proceedings paper
\cite{crrs} and is here systematically developed. In a sense, this
stochastic model takes into account the coupling with a thermal bath
imposing the temperature rather than the energy. Therefore, it
describes a system of stochastically forced vortices in contrast to
the usual point vortex model describing an isolated system. At the
present time, it is not clear how such a thermal bath can be realized
in nature.  However, this stochastic model is
well-defined mathematically and constitutes {\it by definition} the
dynamical model of point vortices associated with  the canonical
ensemble.  Therefore, its study is interesting in its own right since
it provides the out-of-equilibrium version of the situation described
by the Gibbs canonical measure at equilibrium. As we said previously,
it is of conceptual interest to compare the microcanonical and the
canonical dynamics and study them in parallel. On the other hand, by a
proper reinterpretation of the parameters, this stochastic model is
isomorphic to the overdamped dynamics of Brownian particles with
long-range interactions such as self-gravitating Brownian particles
\cite{crs}, bacterial populations experiencing chemotaxis
\cite{ks}, Debye-H\"uckel electrolytes \cite{dh} or electric charges moving in a strong vertical magnetic field  (see Appendix \ref{sec_gcp}). Therefore, 
it provides an interesting model of random walkers in long-range
interactions and enters in the general class of stochastic processes
studied in \cite{hb1,hb2}. In addition, this model allows for many
generalizations with respect to previous works: (i) Point vortices can
have positive and negative circulations. This generalizes the studies
of self-gravitating Brownian particles and bacterial populations where
the mass of the particles is always positive.  (ii) Since point
vortices have no inertia, the temperature can be either positive or
negative. At negative temperatures, point vortices of the same sign
have the tendency to attract each other (like stars in a galaxy
\footnote{The numerous analogies between the statistical
mechanics of 2D point vortices at negative temperatures and the
statistical mechanics of stellar systems have been described in \cite{hou}.}) and at positive
temperatures they have the tendency to repel each other (like
electric charges in a plasma). Therefore, by changing the sign of the
temperature, we can study either attractive or repulsive
interactions. (iii) The equilibrium states of point vortices in the
canonical ensemble have been studied in detail and present peculiar
features associated with the existence of critical temperatures beyond
which there is no equilibrium \cite{km,caglioti,kiessling,diff}. Our dynamical
model will allow to explore this range of parameters where rich and
interesting behaviors are expected to occur. Therefore, on the
viewpoint of statistical mechanics, this stochastic model of Brownian
vortices has very rich properties and deserves a specific attention.

The paper is organized as follows. In Sec. \ref{sec_single}, we
consider a single species system of Brownian vortices.  Starting from
the Fokker-Planck equation satisfied by the $N$-body distribution
function, we derive the BBGKY-like hierarchy for the reduced
distribution functions (see Sec. \ref{sec_ea}). We close this
hierarchy by considering an expansion of the solutions in powers of
$1/N$ for $N\gg 1$. For $N\rightarrow +\infty$ (see
Sec. \ref{sec_mf}), we derive the mean field Fokker-Planck equation
satisfied by the one-body distribution function (smooth vorticity
field). This is the counterpart of the Vlasov regime in plasma
physics. For spatially homogeneous systems, we derive the equation
satisfied by the two-body correlation function at the order $O(1/N)$
and investigate the stability of the system (see
Sec. \ref{sec_dh}). This is the counterpart of the Debye-H\"uckel
theory in plasma physics. In Sec. \ref{sec_exact}, we derive the
stochastic equation satisfied by the exact vorticity field before
averaging over the noise. If we average this equation over the noise,
we recover the first equation of the BBGKY hierarchy. In
Secs. \ref{sec_newton} and
\ref{sec_yukawa}, we consider specific forms of potentials of
interaction corresponding to the ordinary Newtonian (or Coulombian)
potential in two dimensions and to the Yukawa (or Rossby) potential
appropriate to geophysical flows. For spatially homogeneous systems,
we determine the two-body correlation function and the caloric curve
relating the average energy to the temperature. We specifically
distinguish the case of positive and negative temperatures. In
Sec. \ref{sec_two}, we consider a multi-species gas of Brownian
vortices and focus particularly on the two-species system where the
vortices have positive and negative circulations. We derive the
BBGKY-like hierarchy and consider the Vlasov limit $N\rightarrow
+\infty$ for the one-body distribution functions and the
Debye-H\"uckel approximation (valid at the order $O(1/N)$) for the
two-body correlation functions. Finally, in Sec. \ref{sec_conclusion},
we conclude and discuss some perspectives of our work.

\section{The single species system}
\label{sec_single}

\subsection{The stochastic equations}
\label{sec_sto}

The canonical distribution of a single species system of point
vortices with individual circulations $\gamma$ is
\begin{eqnarray}
\label{sto1}
P_{N}=\frac{1}{Z}e^{-\beta H},
\end{eqnarray}
where $H=\sum_{i<j}\gamma^2 u({\bf r}_{i},{\bf
r}_{j})+\frac{1}{2}\gamma^2 \sum_{i=1}^N v({\bf r}_{i},{\bf r}_{i})$
is the Hamiltonian, $Z(\beta)=\int e^{-\beta H}d{\bf r}_{1}...d{\bf r}_{N}$
the partition function and $\beta=1/T$ the inverse temperature which
can be either positive or negative. The free energy is
$F(\beta)=-(1/\beta)\ln Z(\beta)$. We have used the general Green's function
formulation of Lin \cite{lin}, which is valid in an arbitrary bounded
domain
\footnote{In \cite{pl}, the Green function $u({\bf r}_{i},{\bf r}_{j})$ is
noted $-G({\bf r}_{i},{\bf r}_{j})$ and the function $v({\bf
r}_{i},{\bf r}_{i})$ is noted $-g({\bf r}_{i},{\bf r}_{i})$.}. For
convenience, we shall note $U\equiv \sum_{i<j} u({\bf r}_{i},{\bf
r}_{j})+\frac{1}{2}\sum_{i=1}^N v({\bf r}_{i},{\bf r}_{i})$. We ask:
{\it can we formally construct a dynamical model of point vortices
associated with the canonical ensemble?} The answer is yes. The
dynamical model of point vortices consistent with the Gibbs canonical
distribution (\ref{sto1}) at statistical equilibrium is defined by the $N$
coupled stochastic equations
\begin{equation}
\label{sto2}
\frac{d{\bf r}_{i}}{dt}=-\gamma\, {\bf
z}\times\nabla_{i}U-\mu\gamma^2\nabla_iU+\sqrt{2\nu}{\bf R}_{i}(t),
\end{equation}
where $i=1,...,N$ label the point vortices in the system. These
stochastic equations define our model of two-dimensional Brownian
vortices.  The first term, where ${\bf z}$ is a unit vector
perpendicular to the plane of motion, describes the mutual advection
of the point vortices. The last term is a stochastic velocity where
${\bf R}_i(t)$ is a white noise such that $\langle {\bf
R}_i(t)\rangle={\bf 0}$ and $\langle
R_i^{\alpha}(t)R_j^{\beta}(t')\rangle=\delta_{ij}\delta_{\alpha\beta}\delta(t-t')$,
and $\nu\ge 0$ is a diffusion coefficient which plays the role of a
viscosity (see below). Finally, the middle term is a drift velocity
produced by the system as a whole, where $\mu$ plays the role of a
mobility in Brownian theory. The drift velocity is perpendicular to
the advective velocity. This systematic drift is necessary to
compensate the effect of the stochastic velocity and yield the
canonical distribution at equilibrium. This is similar to the
dynamical friction in Brownian theory.

The model (\ref{sto2}) regroups and generalizes  several models of particles
in interaction already introduced in the literature. For
$\mu=\nu=0$, we recover the usual Hamiltonian model \cite{kirchhoff} of point
vortices \footnote{The
Hamiltonian point vortex system is isomorphic to a two-dimensional
guiding center plasma in which charged rods move across an external
magnetic field ${\bf B}=B{\bf z}$ with the self-consistent ${\bf
E}\times {\bf B}$ drift
\cite{jm}. In this analogy, the circulation $\gamma$ plays the role of the charge $e$ and the stream function $\psi$ the role of the electric potential $\Phi$ (see Appendix \ref{sec_gcp}).}:
\begin{equation}
\label{sto3}
\frac{d{\bf r}_{i}}{dt}=-\gamma\, {\bf z}\times\nabla_{i}U,
\end{equation}
which is associated with the microcanonical ensemble
\cite{onsager,pl,lp,bbgky}. For $\mu=0$ and $\nu> 0$, we recover the
stochastic equations considered by Marchioro \& Pulvirenti \cite{mp}:
\begin{equation}
\label{sto4}
\frac{d{\bf r}_{i}}{dt}=-\gamma\, {\bf
z}\times\nabla_{i}U+\sqrt{2\nu}{\bf R}_{i}(t).
\end{equation}
Finally, without the mutual advection (obtained formally by taking
${\bf z}={\bf 0}$), the stochastic equations
\begin{equation}
\label{sto5}
\frac{d{\bf r}_{i}}{dt}=-\mu\gamma^2\nabla_iU+\sqrt{2\nu}{\bf
R}_{i}(t),
\end{equation}
describe the dynamics of Brownian particles in interaction in an
overdamped limit where $\gamma$ plays the role of the mass $m$ and
$\nu$ plays the role of the diffusion coefficient $D_*$ (in that case,
the model is valid in $d$ dimensions) \cite{hb1,hb2}. In the case of
attractive (e.g., Newtonian) potentials, these stochastic equations
describe, for example, a gas of self-gravitating Brownian particles
\cite{crs,virial2} or bacterial populations experiencing chemotaxis
\cite{ng,cskin,chemod}.

\subsection{Ensembles average}
\label{sec_ea}

Using the same procedure as the one developed in Paper II, we can
readily write down the Fokker-Planck equation  associated with the
stochastic process (\ref{sto2}). The evolution of the $N$-body distribution
function $P_N({\bf r}_1,...,{\bf r}_N,t)$ is given by
\begin{eqnarray}
\label{ea1}
\frac{\partial P_{N}}{\partial t}+\sum_{i=1}^{N}{\bf V}_{i}\cdot
\frac{\partial P_{N}}{\partial {\bf r}_{i}}
=\sum_{i=1}^{N}\frac{\partial}{\partial {\bf r}_{i}}\cdot\left
(\nu\frac{\partial P_{N}}{\partial {\bf r}_{i}}+\mu\gamma^2
P_{N}\frac{\partial U}{\partial {\bf r}_{i}}\right ),
\end{eqnarray}
where ${\bf V}_{i}=-\gamma {\bf z}\times {\partial U}/{\partial {\bf
r}_{i}}$ is the total advective velocity of point vortex $i$. For
$\nu=\mu=0$, we recover the Liouville equation corresponding to a
microcanonical description \cite{bbgky}.  For $\nu >0$ and $\mu\neq
0$, the stationary solution of the Fokker-Planck equation (\ref{ea1}) is
\begin{eqnarray}
\label{ea2}
P_{N}=\frac{1}{Z}e^{-\frac{\mu\gamma^2}{\nu}U}.
\end{eqnarray}
It cancels individually the r.h.s. (Fokker-Planck term) and the l.h.s.
(advective term) of Eq. (\ref{ea1}). If we compare this expression
with the Gibbs canonical distribution (\ref{sto1}), we find that the
inverse temperature is related to the mobility and to the diffusion
coefficient (viscosity) by
\begin{eqnarray}
\label{ea3}
\beta=\frac{\mu}{\nu},
\end{eqnarray}
which is the Einstein relation in the present context. Since the
vortices have no inertia, the temperature can take positive or
negative values. Considering Eq. (\ref{sto2}) with the ordinary
potential of interaction (\ref{newton1}) or (\ref{yukawa1}), we see
that at positive temperatures (implying $\mu> 0$) like-sign vortices
repel each other and at negative temperatures (implying $\mu< 0$)
like-sign vortices attract each other.  Let us introduce the free
energy $F=E-TS$ where $E=\langle H\rangle=\int P_{N}H d{\bf
r}_{1}...d{\bf r}_{N}$ is the average energy and $S=-\int P_{N}\ln
P_{N}d{\bf r}_{1}...d{\bf r}_{N}$ the entropy. We also consider the
Massieu functional \footnote{Since the temperature can be negative in
two-dimensional point vortex dynamics, it is often more convenient to
use the Massieu functional (which is the direct Legendre transform of
the entropy) rather than the usual free energy. By an abuse of
language, we shall often refer to $J$ as the ``free energy''.} $J\equiv
-\beta F=S-\beta E$ given by
\begin{eqnarray}
\label{ea4}
J[P_N]=-\int P_{N}\ln P_{N} d{\bf r}_{1}...d{\bf r}_{N}-\frac{1}{2}\beta \int P_{N}H
d{\bf r}_{1}...d{\bf r}_{N}.
\end{eqnarray} 
We see that the Gibbs state (\ref{sto1}) is a critical point of $J$ with the normalization constraint: $\delta J-\alpha
\int \delta P_{N}d{\bf r}_{1}...d{\bf r}_{N}=0$. Furthemore,
$\delta^{2}J=-(1/2)\int (\delta P_{N})^{2}/P_{N}d{\bf r}_{1}...d{\bf
r}_{N}<0$. Therefore, if the Gibbs state exists (i.e. if the partition
function exists), then it is the only maximum of the free energy
functional $J[P_N]$. Furthermore, inserting the Gibbs state
(\ref{sto1}) in the Massieu functional (\ref{ea4}), we recover the
equilibrium value of the free energy $J(\beta)=\ln Z(\beta)$. On the
other hand, using the Fokker-Planck equation (\ref{ea1}) we find that
\begin{eqnarray}
\label{ea5}
\dot J=\sum_{i=1}^{N}\int \frac{1}{\nu P_{N}}\left (\nu \frac{\partial P_{N}}{\partial {\bf r}_{i}}+\mu\gamma^2 P_{N}\frac{\partial U}{\partial {\bf r}_{i}}\right )^{2}d{\bf r}_{1}...d{\bf r}_{N}.
\end{eqnarray}
Therefore, if $\mu\neq 0$ and $\nu>0$ the Fokker-Planck equation
(\ref{ea1}) satisfies an $H$-theorem appropriate to the canonical
ensemble: $\dot J\ge 0$ and $\dot J=0$ iff $P_N$ is the canonical
distribution (\ref{sto1}). The free energy (\ref{ea4}) is the Lyapunov
functional of the Fokker-Planck equation (\ref{ea1}). From Lyapunov's
direct method, we conclude that if $J$ is bounded from above (or,
equivalently, if the partition function exists), the Fokker-Planck
equation (\ref{ea1}) will relax, for $t\rightarrow +\infty$, towards
the Gibbs distribution (\ref{sto1}).

Proceeding as in Paper II, we can easily obtain the exact hierarchy
of equations satisfied by the reduced distributions $P_j({\bf
r}_1,...,{\bf r}_j,t)$. To that purpose, we first note that  the
stochastic equation (\ref{sto2}) takes the form of Eq. (II-146) if we make the
substitutions $m\rightarrow \gamma$, $D_*\rightarrow \nu$ and
\begin{equation}
\label{ea6}
\frac{\partial u_{ij}}{\partial {\bf r}_i}\rightarrow
\frac{\partial'{u}_{ij}}{\partial {\bf r}_i}\equiv
\frac{1}{\mu\gamma}{\bf z}\times \frac{\partial u_{ij}}{\partial {\bf
r}_i}+\frac{\partial u_{ij}}{\partial {\bf r}_i},
\end{equation}
where $\partial'u_{ij}/\partial {\bf r}_i$ is a convenient
notation. We also need to take into account the term
$\frac{1}{2}\sum_{i=1}^N v({\bf r}_{i},{\bf r}_{i})$ in the
Hamiltonian. With these substitutions, we can immediately transpose
the results of Paper II to the present context. The equations for
the reduced distributions of order $j=1,...,N$ are given by
\begin{equation}
\label{ea7}  {\partial P_{j}\over\partial t}=\sum_{i=1}^{j}\frac{\partial}{\partial {\bf r}_{i}}\biggl \lbrack \nu \frac{\partial P_{j}}{\partial {\bf r}_{i}}+\mu\gamma^2\sum_{k=1,k\neq i}^{j}P_{j}\frac{\partial' u_{ik}}{\partial {\bf r}_{i}}+\mu\gamma^2(N-j)\int P_{j+1}\frac{\partial' u_{i,j+1}}{\partial {\bf r}_{i}}d{\bf r}_{j+1}+\frac{\mu\gamma^2}{2} P_{j}\frac{\partial'}{\partial {\bf r}_{i}}v({\bf r}_{i},{\bf r}_{i})\biggr \rbrack
\end{equation}
and they form the BBGKY-like hierarchy associated with the
Fokker-Planck equation (\ref{ea1}). The equilibrium distributions are obtained
by cancelling the terms in brackets and by replacing $\partial'$ by
$\partial$ (since the steady state cancels the advective
term). They satisfy the equilibrium  BBGKY-like hierarchy
\begin{equation}
\label{ea7b}  \frac{\partial P_{j}}{\partial {\bf r}_{1}}=-\beta\gamma^2\sum_{k=2}^{j}P_{j}\frac{\partial u_{1,k}}{\partial {\bf r}_{1}}-\beta\gamma^2(N-j)\int P_{j+1}\frac{\partial u_{1,j+1}}{\partial {\bf r}_{1}}d{\bf r}_{j+1}-\frac{\beta\gamma^2}{2} P_{j}\frac{\partial}{\partial {\bf r}_{1}}v({\bf r}_{1},{\bf r}_{1}).
\end{equation}
They can also be obtained directly from the Gibbs canonical
distribution (\ref{sto1}) by writing
\begin{equation}
\label{ea8}  {\partial P_{N}\over\partial {\bf r}_1}=-\beta\gamma^2
P_N \frac{\partial U}{\partial {\bf r}_1},
\end{equation}
and integrating over ${\bf r}_{j+1}...{\bf r}_N$. 
The exact first two
equations of the BBGKY-like hierarchy are
\begin{equation}
\label{ea9}  {\partial P_{1}\over\partial t}=
{\partial\over\partial {\bf r}_{1}}\biggl\lbrack \nu {\partial
P_{1}\over\partial {\bf r}_{1}}+\mu \gamma^{2}(N-1)\int
 {\partial' {u}_{12}\over\partial {\bf r}_{1}}P_{2}
d{\bf r}_{2}+\frac{1}{2}\mu\gamma^2 P_1 \frac{\partial'}{\partial{\bf
r}_1}v({\bf r}_1,{\bf r}_1)\biggr\rbrack,
\end{equation}
\begin{eqnarray}
\label{ea10}{\partial P_{2}\over\partial t}={\partial\over\partial
{\bf r}_{1}}\biggl\lbrack \nu {\partial P_{2}\over\partial {\bf
r}_{1}}+\mu \gamma^{2} P_{2}{\partial' {u}_{12}\over\partial {\bf
r}_{1}}+(N-2)\mu \gamma^{2} \int {\partial' {u}_{13}\over\partial
{\bf r}_{1}}P_{3} d{\bf r}_{3}+\frac{1}{2}\mu\gamma^2 P_2
\frac{\partial'}{\partial{\bf r}_1}v({\bf r}_1,{\bf
r}_1)\biggr\rbrack\nonumber\\
+(1\leftrightarrow 2).
\end{eqnarray}
The next step is to decompose the reduced distributions in the form of
a Mayer expansion by introducing the cumulants (see Eqs (I-14) and
(I-15) of Paper I). Inserting the decomposition (I-14) in
Eq. (\ref{ea9}), we obtain
\begin{eqnarray}
\label{ea11}  {\partial P_{1}\over\partial t}=
{\partial\over\partial {\bf r}_{1}}\biggl\lbrack \nu {\partial
P_{1}\over\partial {\bf r}_{1}}+\mu \gamma^{2}(N-1)\int
 {\partial' {u}_{12}\over\partial {\bf r}_{1}}P_{1}({\bf r}_{1})P_{1}({\bf r}_{2})
d{\bf r}_{2}\nonumber\\
+\mu \gamma^{2}(N-1)\int
 {\partial' {u}_{12}\over\partial {\bf r}_{1}}P'_{2}({\bf r}_{1},{\bf r}_{2})
d{\bf r}_{2}+\frac{1}{2}\mu\gamma^2 P_1 \frac{\partial'}{\partial{\bf
r}_1}v({\bf r}_1,{\bf r}_1)\biggr\rbrack,
\end{eqnarray}
where $P_2'$ is the two-body correlation function. Next, inserting
the decomposition (I-14), (I-15) in Eq. (\ref{ea10}) and using Eq.
(\ref{ea11}) to simplify some terms, we get
\begin{eqnarray}
\label{ea12}{\partial P_{2}'\over\partial t}={\partial\over\partial
{\bf r}_{1}}\biggl\lbrack \nu {\partial P_{2}'\over\partial {\bf
r}_{1}}+\mu \gamma^{2} P_{1}({\bf r}_{1}) P_{1}({\bf
r}_{2}){\partial' {u}_{12}\over\partial {\bf r}_{1}}+\mu \gamma^{2}
P_{2}'({\bf r}_{1},{\bf r}_{2}){\partial' {u}_{12}\over\partial {\bf
r}_{1}}
\nonumber\\
-\mu \gamma^{2}\int {\partial' {u}_{13}\over\partial {\bf
r}_{1}}P_{1}({\bf r}_{1}) P_{1}({\bf r}_{2})P_{1}({\bf r}_{3})d{\bf
r}_{3}+(N-2)\mu  \gamma^{2} \int{\partial' {u}_{13}\over\partial
{\bf r}_{1}} P_{2}'({\bf r}_{1},{\bf r}_{2})
P_{1}({\bf r}_{3})d{\bf r}_{3}\nonumber\\
-\mu \gamma^{2}\int {\partial' {u}_{13}\over\partial {\bf
r}_{1}}P_{2}' ({\bf r}_{1},{\bf r}_{3})P_{1}({\bf r}_{2})d{\bf
r}_{3}+ (N-2)\mu \gamma^{2} \int {\partial' {u}_{13}\over\partial
{\bf r}_{1}} P_{2}'({\bf r}_{2},{\bf r}_{3})
P_{1}({\bf r}_{1})d{\bf r}_{3}\nonumber\\
+(N-2)\mu  \gamma^{2}\int {\partial' {u}_{13}\over\partial {\bf
r}_{1}}P_{3}'({\bf r}_{1},{\bf r}_{2},{\bf r}_{3})d{\bf
r}_{3}+\frac{1}{2}\mu\gamma^2 P_2' \frac{\partial'}{\partial{\bf
r}_1}v({\bf r}_1,{\bf r}_1)\bigg\rbrack+(1\leftrightarrow 2),
\end{eqnarray}
where $P_3'$ is the three-body correlation function. These equations
are exact for all $N$, but the hierarchy is not closed.

We now consider the thermodynamic limit $N\rightarrow +\infty$ in such
a way that the normalized inverse temperature $\eta\equiv \beta
N\gamma^2$ remains of order unity. This corresponds to a regime of
{\it weak coupling} since the ``plasma parameter'' $\beta\gamma^2
u\sim 1/N\rightarrow 0$. It is convenient to renormalize the
parameters in such a way that the individual circulations scale like
$\gamma\sim 1/N$ (so that the total circulation $\Gamma\sim N\gamma$
is of order unity), the inverse temperature scales like $\beta\sim N$
and the domain area scales like $V\sim 1$.  We also assume that
$\nu\sim 1$ so that the diffusive timescale is of order unity. The
dynamical time is also of order unity since $t_D\sim 1/\omega\sim
V/\Gamma\sim 1$. With these scalings, we have $P_{1}\sim 1$, $|{\bf
r}|\sim 1$, $u\sim 1$ and $\mu\sim\beta\sim N$. Now, by considering
the scaling of the terms appearing in the different equations of the
BBGKY-like hierarchy, we find that the cumulants scale like
$P_{j}'\sim 1/N^{j-1}$. In particular, $P_{2}'\sim 1/N$ and
$P_{3}'\sim 1/N^{2}$. Therefore, up to the order $1/N$, the first two
equations of the BBGKY-like hierarchy take the form
\begin{eqnarray}
\label{ea13}  {\partial P_{1}\over\partial t}=
{\partial\over\partial {\bf r}_{1}}\biggl\lbrack \nu {\partial
P_{1}\over\partial {\bf r}_{1}}+\mu \gamma^{2}(N-1)\int
 {\partial' {u}_{12}\over\partial {\bf r}_{1}}P_{1}({\bf r}_{1})P_{1}({\bf r}_{2})
d{\bf r}_{2}\nonumber\\
+\mu \gamma^{2}N\int
 {\partial' {u}_{12}\over\partial {\bf r}_{1}}P'_{2}({\bf r}_{1},{\bf r}_{2})
d{\bf r}_{2}+\frac{1}{2}\mu\gamma^2 P_1 \frac{\partial'}{\partial{\bf
r}_1}v({\bf r}_1,{\bf r}_1)\biggr\rbrack,
\end{eqnarray}
\begin{eqnarray}
\label{ea14}{\partial P_{2}'\over\partial t}={\partial\over\partial
{\bf r}_{1}}\biggl\lbrack \nu {\partial P_{2}'\over\partial {\bf
r}_{1}}+\mu \gamma^{2} P_{1}({\bf r}_{1}) P_{1}({\bf
r}_{2}){\partial' {u}_{12}\over\partial {\bf r}_{1}}
\nonumber\\
-\mu \gamma^{2}\int {\partial' {u}_{13}\over\partial {\bf r}_{1}}
P_{1}({\bf r}_{1})P_{1}({\bf r}_{2})P_{1}({\bf r}_{3})d{\bf r}_{3}
+N\mu  \gamma^{2}\int{\partial' {u}_{13}\over\partial {\bf r}_{1}}
P_{2}'({\bf r}_{1},{\bf r}_{2}) P_{1}({\bf r}_{3})d{\bf r}_{3}\nonumber\\
+N\mu \gamma^{2} \int {\partial' {u}_{13}\over\partial {\bf r}_{1}}
P_{2}'({\bf r}_{2},{\bf r}_{3})P_{1}({\bf r}_{1})d{\bf
r}_{3}\biggr\rbrack+(1\leftrightarrow 2).
\end{eqnarray}
The hierarchy is now closed because the three-body correlation
function can be neglected at the order $O(1/N)$. 

On the other hand, the average energy
\begin{eqnarray}
\label{ea15} E=\langle H\rangle=\sum_{i<j}\int \gamma^2 u_{ij}P_N
d{\bf r}_1...d{\bf r}_N+\frac{1}{2}\sum_{i=1}^N\int \gamma^2
v_{ii}P_N d{\bf r}_1...d{\bf r}_N,
\end{eqnarray}
can be written
\begin{eqnarray}
\label{ea16} E=\frac{1}{2}N(N-1)\gamma^2\int P_2({\bf r}_1,{\bf r}_2)
u_{12}d{\bf r}_1 d{\bf r}_2+ \frac{N}{2}\gamma^2\int P_1({\bf r}_1)
v({\bf r}_1,{\bf r}_1) d{\bf r}_1.
\end{eqnarray}
Substituting the Mayer decomposition (I-14) in the previous equation,
and introducing the smooth vorticity field $\omega=N\gamma P_1$ and the stream function
\begin{eqnarray}
\label{mf3}
\psi({\bf r},t)=\int u({\bf r},{\bf r}')\omega({\bf r}',t)d{\bf r}',
\end{eqnarray}
we obtain
\begin{eqnarray}
\label{ea17} E=\frac{1}{2}\int \omega\psi d{\bf r}-N\gamma^2\int
P_1({\bf r}_1)u_{12}P_1({\bf r}_2) d{\bf r}_1d{\bf
r}_2\nonumber\\
+\frac{1}{2}N(N-1)\gamma^2\int P_2'({\bf r}_1,{\bf
r}_2) u_{12}d{\bf r}_1 d{\bf r}_2+ \frac{N}{2}\gamma^2\int P_1({\bf
r}_1) v({\bf r}_1,{\bf r}_1) d{\bf r}_1.
\end{eqnarray}

\subsection{The mean field approximation}
\label{sec_mf}

For $N\rightarrow +\infty$, we can neglect the two-body correlation
function $P_2'\sim 1/N$. In that case, the two-body distribution
function factorizes in two one-body distribution functions:
\begin{eqnarray}
\label{mf1}
P_2({\bf r}_1,{\bf r}_2,t)=P_1({\bf r}_1,t)P_1({\bf r}_2,t)+O(1/N).
\end{eqnarray}
More generally, for $N\rightarrow +\infty$, we have the factorization
\begin{eqnarray}
\label{mf1gen}
P_N({\bf r}_1,...,{\bf r}_{N},t)=\prod_{i=1}^{N}P_1({\bf r}_i,t).
\end{eqnarray}
This is the equivalent of the mean field approximation (valid in the
Vlasov regime) in plasma physics or stellar dynamics. The mean field
approximation is exact at the thermodynamic limit $N\rightarrow
+\infty$. Introducing the smooth vorticity field $\omega=N\gamma P_1$
and the stream function (\ref{mf3}), the first equation (\ref{ea13})
of the BBGKY-like hierarchy reduces to the mean field Fokker-Planck
equation
\footnote{To avoid confusions, it should be emphasized that the mean
field Fokker-Planck equation (\ref{mf2}) has a completely different
origin and interpretation from the Fokker-Planck equation (140)
derived in \cite{bbgky}. In \cite{bbgky}, the Fokker-Planck equation
(140) describes the relaxation of a test vortex (tagged particle) in a
{\it fixed} distribution of field vortices at statistical equilibrium
(thus $\psi({\bf r})$ is a given function independent on time).  Here,
the mean field Fokker-Planck equation (\ref{mf2}) describes the
evolution of a Brownian system of point vortices as a whole for an
assumed microscopic dynamics of the form (\ref{sto2}) (thus $\psi({\bf
r},t)$ is a function of time produced self-consistently by the
distribution of point vortices according to Eq. (\ref{mf3})). The mean
field Fokker-Planck equation (\ref{mf2}) is also physically different
from the relaxation equation (12) with a time dependent temperature
$\beta(t)$ introduced by Robert
\& Sommeria
\cite{rsmepp} to phenomenologically describe the violent relaxation of
the 2D Euler equation on a coarse-grained scale.}
\begin{eqnarray}
\label{mf2}
\frac{\partial \omega}{\partial t}+{\bf u}\cdot \nabla \omega=\nu
\nabla \cdot (\nabla \omega+\beta\gamma \omega\nabla \psi)\equiv
-\nabla\cdot {\bf J},
\end{eqnarray}
where ${\bf u}=-{\bf z}\times \nabla\psi$ is the mean field
velocity. The mean field Fokker-Planck equation (\ref{mf2}) conserves the
circulation $\Gamma=N\gamma=\int \omega d{\bf r}$, proportional to the
vortex number. On the other hand, its steady solution is
the mean field Boltzmann distribution
\begin{eqnarray}
\label{mf4}
\omega({\bf r})=Ae^{-\beta \gamma\psi({\bf r})}.
\end{eqnarray}
This distribution cancels individually the advective term (l.h.s.) and
the Fokker-Planck term (r.h.s.) of Eq. (\ref{mf2}). The mean field
Boltzmann distribution can also be obtained from the equilibrium
BBGKY-like hierarchy in the limit $N\rightarrow +\infty$ (see Paper
I). Substituting Eq. (\ref{mf4}) in Eq. (\ref{mf3}), we obtain an
integro-differential equation $\nabla\ln\omega=-\beta\gamma\int\nabla
u({\bf r},{\bf r}')\omega({\bf r}')\, d{\bf r}'$ determining the
vorticity field at statistical equilibrium.

For $N\rightarrow +\infty$, using
Eq. (\ref{mf1gen}), the Boltzmann entropy $S=-\int P_{N}\ln P_{N}
d{\bf r}_{1}...d{\bf r}_{N}$ reduces to
\begin{eqnarray}
\label{mf5}
S=-N\int P_1({\bf r},t)\, \ln P_1({\bf r},t)\, d{\bf r}.
\end{eqnarray}
It can be written (up to an additive constant)
\begin{eqnarray}
\label{mf6}
S[\omega]=-\int \frac{\omega}{\gamma} \ln \frac{\omega}{\gamma}\, d{\bf r}.
\end{eqnarray}
This expression of the Boltzmann entropy can also be obtained from a
classical combinatorial analysis \cite{jm,cl}.  On the other hand, for
$N\rightarrow +\infty$, we obtain from Eq. (\ref{ea17}) the mean field
energy
\begin{eqnarray}
\label{mf7} E[\omega]=\frac{1}{2}\int \omega\psi d{\bf r}.
\end{eqnarray}
The mean field free energy (more precisely the mean field Massieu
function) is $J[\omega]=S[\omega]-\beta E[\omega]$ where $S$ and $E$
are given by Eqs. (\ref{mf6}) and (\ref{mf7}). It can be obtained from
the free energy (\ref{ea4}) by using the mean field approximation
(\ref{mf1gen}) valid for $N\rightarrow +\infty$ \cite{hb2}.  The
kinetic equation (\ref{mf2}) can then be rewritten in the form
\begin{eqnarray}
\label{mf8} \frac{\partial\omega({\bf r},t)}{\partial t}-\left ({\bf
z}\times\nabla\frac{\delta E\lbrack\omega\rbrack}{\delta\omega({\bf
r},t)}\right )\cdot\nabla\omega({\bf
r},t)=-\nabla\cdot\left\lbrack\nu\gamma\omega({\bf
r},t)\nabla\frac{\delta J\lbrack\omega\rbrack}{\delta\omega({\bf
r},t)}\right\rbrack.
\end{eqnarray}
The steady state corresponds to a uniform value of
$\alpha={\delta J}/{\delta\omega({\bf r})}$ leading to the Boltzmann distribution (\ref{mf4}). On the other hand, it is
straightforward to establish that
\begin{eqnarray}
\label{mf9}
\dot J=\int \frac{{\bf J}^2}{\nu\gamma\omega}d{\bf r}=\int
\nu\gamma\omega \left (\nabla\frac{\delta J}{\delta\omega}\right)^2
d{\bf r}\ge 0.
\end{eqnarray}
Therefore, if $\mu\neq 0$ and $\nu>0$ the mean field Fokker-Planck
equation (\ref{mf2}) satisfies an $H$-theorem appropriate to the
canonical ensemble: $\dot J\ge 0$ and $\dot J=0$ iff $\omega$ is the
mean field Boltzmann distribution (\ref{mf4}). The free energy
$J[\omega]$ is the Lyapunov functional of the Fokker-Planck equation
(\ref{mf2}). The Boltzmann distribution (\ref{mf4}) is a critical
point of $J$ at fixed $\Gamma$ (cancelling the first variations) and
it is linearly dynamically stable iff it is a (local) maximum of $J$
at fixed $\Gamma$. If $J$ is bounded from above, we conclude from
Lyapunov's direct method, that the mean field Fokker-Planck equation
(\ref{mf2}) will reach, for $t\rightarrow +\infty$, a (local) maximum
of $J$ at fixed circulation. If several local maxima exist, the
selection of the maximum will depend on a complicated notion of basin
of attraction. If there is no maximum of free energy at fixed
circulation, the system will have a peculiar behavior and will
generate singularities associated with vortex collapse (see
Secs. \ref{sec_newton} and \ref{sec_yukawa}). We note that the
relaxation equation (\ref{mf2}) can be obtained from a maximum entropy
production principle (MEPP) by maximizing the production of free
energy $\dot J$ at fixed circulation (and other physical constraints)
\cite{gen,nfp}.

Finally, we note that the corresponding equation for the velocity field ${\bf
u}=-{\bf z}\times \nabla\psi$ is
\begin{eqnarray}
\label{mf10}
\frac{\partial {\bf u}}{\partial t}+({\bf u}\cdot \nabla){\bf u}=-\frac{1}{\rho}\nabla p+\nu\Delta{\bf u}-\nu\beta\gamma\omega{\bf u}, 
\end{eqnarray}
where $p({\bf r},t)$ is a pressure field. Indeed, taking the curl of
this equation and using $\omega{\bf z}=\nabla\times {\bf u}$, we
recover Eq. (\ref{mf2}) [the intermediate steps of the calculation make
use of the identity of vector analysis $\Delta{\bf u}=\nabla
(\nabla\cdot {\bf u})-\nabla\times (\nabla\times {\bf u})$ which
reduces to $\Delta {\bf u}=-\nabla\times (\omega{\bf z})={\bf z}\times
\nabla\omega$ for an incompressible velocity field ${\bf u}=-{\bf z}\times \nabla\psi$. On the other hand,
$\nabla\times ({\bf z}\times \nabla\omega)=\Delta\omega{\bf z}$ and
$\nabla\times (\omega{\bf u})=\omega \nabla\times{\bf
u}+\nabla\omega\times {\bf u}=-\nabla\cdot (\omega\nabla\psi){\bf z}$;
finally, the transformation of the advective term is
standard]. Interestingly, we note that the systematic drift of the
point vortices $-\mu\gamma\nabla\psi$ in Eq. (\ref{mf2}) corresponds
to a friction force $-\mu\gamma\omega{\bf u}$ in the equation
(\ref{mf10}) for the velocity. This ``duality'' is a particularity of
the Brownian point vortex system which has no counterpart in systems
of material particles.

\subsection{The two-body correlation function}
\label{sec_dh}

The evolution of the two-body correlation function at the order $1/N$
is determined by the second equation (\ref{ea14}) of the BBGKY
hierarchy.  Let us consider a spatially homogeneous distribution of
point vortices (to leading order) so that $P_1({\bf
r},t)=P_0+\hat{P}({\bf r},t)$ where $\hat{P}$ is of order $1/N$. Here,
we consider an infinite domain so that $u({\bf
r},{\bf r}')=u(|{\bf r}-{\bf r}'|)$. Writing $P_2'({\bf r},{\bf
r}',t)=P_0^2 h(|{\bf r}-{\bf r}'|,t)$, the second equation (\ref{ea14})
of the BBGKY hierarchy equation simplifies into
\begin{eqnarray}
\label{dh1}{\partial h\over\partial t}=2\nu \Delta \left\lbrack
h+\beta \gamma^2 u+\beta n\gamma^2 \int h({\bf y},t)u(|{\bf x}-{\bf
y}|)d{\bf y}\right\rbrack,
\end{eqnarray}
where ${\bf x}={\bf r}-{\bf r}'$. The steady distribution of this
equation satisfies an equation of the form
\begin{eqnarray}
\label{dh2} \frac{\partial h}{\partial {\bf x}}=-\beta \gamma^2
\frac{\partial u}{\partial {\bf x}}-\beta n\gamma^2 \int h({\bf
y},t)\frac{\partial u}{\partial {\bf x}}(|{\bf x}-{\bf y}|)d{\bf y}.
\end{eqnarray}
Noting that the integral is a convolution, this integro-differential
equation is easily solved in Fourier space leading to
\begin{eqnarray}
\label{dh3} \hat{h}_{eq}({\bf k})=\frac{-\beta\gamma^2 \hat{u}({k})}{1+(2\pi)^2 \beta n\gamma^2 \hat{u}({k})}.
\end{eqnarray}
The dynamical equation (\ref{dh1}) can be solved similarly and the
temporal evolution of the two-body correlation function in Fourier
space with initial condition $\hat{h}({\bf k},0)=0$ is found to be
\begin{equation} \label{dh4}
\hat{h}({\bf k},t)=\hat{h}_{eq}({k})\left\lbrace 1 -e^{-2\nu
k^{2}\left\lbrack 1+(2\pi)^{2}\beta n
\gamma^{2}\hat{u}(k)\right\rbrack t}\right\rbrace.
\end{equation}
From this equation, we see  that the homogeneous equilibrium distribution is
stable iff
\begin{eqnarray}
\label{dh5} {1+(2\pi)^2 \beta n\gamma^2 \hat{u}({k})}>0,
\end{eqnarray}
for any wave vector ${\bf k}$ (otherwise, the wave vectors that do not
satisfy this relation correspond to unstable modes that grow
exponentially rapidly with time). This inequality usually determines a
critical temperature $\beta_0$ (similar to a spinodal point) beyond
which the homogeneous system becomes unstable \cite{hb1}. The
stability criterion (\ref{dh5}) can also be obtained by studying the
linear dynamical stability of a spatially homogeneous steady solution
of the mean field Fokker-Planck equation (\ref{mf2}) with
$\omega=N\gamma P_0$ and ${\bf u}={\bf 0}$. Indeed, decomposing  the
perturbation in normal modes  as $\delta
\omega\sim e^{i({\bf k}\cdot {\bf r}-\sigma t)}$, the dispersion
relation obtained by linearizing  the Fokker-Planck equation (\ref{mf2}) is
$i\sigma =\nu k^2
\left\lbrack 1+(2\pi)^2 \beta n\gamma^2 \hat{u}({k})\right\rbrack$. The system is stable iff $i\sigma>0$ for all
${\bf k}$ corresponding to the inequality (\ref{dh5}). This inequality
is also implied by Eq. (\ref{id5}) of Appendix \ref{sec_id} and by the
condition that the equilibrium state must be a maximum of $J[\omega]$
at fixed $\Gamma[\omega]$ (see Sec. 4.4. of \cite{hb1}).

\subsection{The exact vorticity field}
\label{sec_exact}

In this section, we shall derive the stochastic equation satisfied
by the exact vorticity field before averaging over the noise. To
simplify the expressions, we shall work in an infinite domain where
$u_{ij}=u(|{\bf r}_i-{\bf r}_j|)$ and $v_{ii}=0$. The generalization
of our results to a bounded domain is obtained by replacing $u(|{\bf
r}_i-{\bf r}_j|)$ by $u({\bf r}_i,{\bf r}_j)$ and by adding a term
$(1/2)\gamma^2 \nabla v({\bf r},{\bf r})$ in the velocity field. The
exact vorticity field is expressed as a sum of Dirac distributions
in the form
\begin{equation}
\label{exact1}
\omega_{d}({\bf r},t)=\gamma \sum_{i}^{N}\delta({\bf r}-{\bf
r}_{i}(t)).
\end{equation}
Adapting the calculations of \cite{kk,dean,mt,ar,prep,chemod} to the present context, we find that the exact vorticity field is
solution of the stochastic equation
\begin{eqnarray}
\label{exact2}
\frac{\partial\omega_d}{\partial t}({\bf r},t)-{\bf z}\times \nabla
\cdot \left (\omega_d({\bf r},t) \nabla \int  \omega_{d}({\bf
r}',t)u(|{\bf r}-{\bf r}'|) d{\bf r}' \right )
=\nu\Delta\omega_{d}({\bf r},t)\nonumber\\
+\mu\gamma\nabla \cdot \left (\omega_{d}({\bf r},t)\nabla\int
\omega_{d}({\bf r}',t)u(|{\bf r}-{\bf r}'|)d{\bf r}'\right ) +\nabla
\cdot \left (\sqrt{2\nu\gamma\omega_{d}({\bf r},t)}{\bf R}({\bf
r},t)\right ),\qquad
\end{eqnarray}
where ${\bf R}({\bf r},t)$ is a Gaussian random field such that
$\langle {\bf R}({\bf r},t)\rangle={\bf 0}$ and $\langle
R^{\alpha}({\bf r},t)R^{\beta}({\bf
r}',t')\rangle=\delta_{\alpha\beta}\delta({\bf r}-{\bf
r}')\delta(t-t')$. Introducing the exact stream function
\begin{eqnarray}
\label{exact3}
\psi_{d}({\bf r},t)= \int  \omega_{d}({\bf r}',t)u(|{\bf r}-{\bf r}'|)\, d{\bf r}',
\end{eqnarray}
and the exact velocity field ${\bf u}_{d}=-{\bf z}\times
\nabla\psi_{d}$, the stochastic equation (\ref{exact2}) can be rewritten in the
form
\begin{eqnarray}
\label{exact4}
\frac{\partial\omega_d}{\partial t}+{\bf u}_{d}\cdot
\nabla\omega_{d}=\nu\Delta\omega_{d} +\nabla\cdot (\mu\gamma
\omega_{d}\nabla\psi_{d})+\nabla\cdot
(\sqrt{2\nu\gamma\omega_{d}}{\bf R}).
\end{eqnarray}
Let us consider some particular cases. For $\nu=\mu=0$, we find that
the exact vorticity field is solution of the 2D Euler equation
\begin{eqnarray}
\label{exact5}
\frac{\partial\omega_d}{\partial t}+{\bf u}_{d}\cdot \nabla\omega_{d}=0.
\end{eqnarray}
This is the counterpart of the Klimontovich equation in plasma
physics. On the other hand, for the model considered by Marchioro \&
Pulvirenti \cite{mp} where $\mu=0$ and $\nu> 0$, the exact
vorticity field is solution of the stochastic equation
\begin{eqnarray}
\label{exact6}
\frac{\partial\omega_d}{\partial t}+{\bf u}_{d}\cdot \nabla\omega_{d}=\nu\Delta\omega_{d}+\nabla\cdot (\sqrt{2\nu\gamma\omega_{d}}{\bf R}).
\end{eqnarray}
This equation was not given in \cite{mp}. Introducing the exact free energy
functional
\begin{eqnarray}
\label{exact7}
{\cal J}\equiv {\cal S}-\beta{\cal H}=-\int \frac{\omega_d({\bf
r},t)}{\gamma}\ln \frac{\omega_d({\bf r},t)}{\gamma} d{\bf
r}-\frac{1}{2}\beta\int \omega_d({\bf r},t)u(|{\bf r}-{\bf
r}'|)\omega_d({\bf r}',t)d{\bf r}d{\bf r}',
\end{eqnarray}
the stochastic equation (\ref{exact4}) can be rewritten in the form
\begin{eqnarray}
\label{exact8} \frac{\partial\omega_d({\bf r},t)}{\partial t}-\left
({\bf z}\times\nabla\frac{\delta {\cal
H}\lbrack\omega_d\rbrack}{\delta\omega_d({\bf r},t)}\right
)\cdot\nabla\omega_d({\bf r},t)\nonumber\\
=-\nabla\cdot\left\lbrack
\nu\gamma\omega_d({\bf r},t)\nabla\frac{\delta {\cal
J}\lbrack\omega_d\rbrack}{\delta\omega_d({\bf
r},t)}\right\rbrack+\nabla\cdot \left\lbrack\sqrt{2\nu\gamma\omega_d({\bf r},t)}{\bf R}({\bf
r},t)\right\rbrack.
\end{eqnarray}
It can be shown \cite{prep} that this stochastic equation reproduces
the equilibrium two-body correlation function $\langle\delta\omega({\bf r})\delta\omega({\bf r}')\rangle$  that is related to the function $h({\bf r}-{\bf r}')$ determined  in Sec. \ref{sec_dh} according
to the identity (\ref{id3}). Equation (\ref{exact8}) can be viewed as
a Langevin equation for the field $\omega_{d}({\bf r},t)$.  Using
standard methods \cite{ar}, the corresponding Fokker-Planck equation
for the probability $W[\omega_{d},t]$ of the field $\omega_d$ is
\begin{eqnarray}
\label{exact9}
\frac{\partial W\lbrack \omega_d,t\rbrack}{\partial t}+\int
\frac{\delta}{\delta\omega_d({\bf r},t)}\left\lbrace \left ({\bf
z}\times\nabla\frac{\delta{\cal
H}\lbrack\omega_d\rbrack}{\delta\omega_d({\bf r},t)}\right
)\cdot\nabla\omega_d({\bf r},t) W\lbrack\omega_d,t\rbrack\right\rbrace
d{\bf r}\nonumber\\
=-\nu\gamma\int\frac{\delta}{\delta\omega_d({\bf r},t)}\left\lbrace
\nabla\cdot \omega_d({\bf r},t)\nabla\left\lbrack
\frac{\delta}{\delta\omega_d({\bf r},t)}-\frac{\delta {\cal
J}\lbrack\omega_d\rbrack}{\delta\omega_d({\bf r},t)}\right\rbrack
W\lbrack\omega_d,t\rbrack\right\rbrace d{\bf r},
\end{eqnarray}
and the equilibrium distribution is $W[\omega_{d}]\propto e^{{\cal
J}[\omega_{d}]-\alpha\omega_{d}}$.

If we now average Eq. (\ref{exact2}) over the noise, we find that the
evolution of the smooth vorticity field $\omega({\bf r},t)=\langle
\omega_d
\rangle$ is governed by an equation of the form
\begin{eqnarray}
\label{exact10}
\frac{\partial\omega}{\partial t}({\bf r},t)- {\bf z}\times \nabla \cdot \int
\langle \omega_d({\bf r},t)  \omega_{d}({\bf r}',t)\rangle \nabla u(|{\bf r}-{\bf r}'|)d{\bf r}'\nonumber\\
=\nu\Delta\omega({\bf r},t)+\mu\gamma\nabla \cdot \int  \langle
\omega_{d}({\bf r},t)\omega_{d}({\bf r}',t)\rangle \nabla u(|{\bf
r}-{\bf r}'|)d{\bf r}'.
\end{eqnarray}
This equation is equivalent to Eq. (\ref{ea9}) giving the exact
evolution of the one-body distribution function. Indeed, using
$\omega({\bf r},t)=N\gamma P_1({\bf r},t)$ and the identity (see
Appendix \ref{sec_id}):
\begin{eqnarray}
\label{exact11}
\langle \omega_d({\bf r},t)  \omega_{d}({\bf r}',t)\rangle=N\gamma^2 P_{1}({\bf r},t)\delta({\bf r}-{\bf r}')
+N(N-1)\gamma^2 P_{2}({\bf r},{\bf r}',t),
\end{eqnarray}
we find that Eqs. (\ref{exact10}) and (\ref{ea9})
coincide. Furthermore, if we make the mean field approximation
$\langle \omega_d({\bf r},t)
\omega_{d}({\bf r}',t)\rangle=\omega({\bf r},t)\omega({\bf r}',t)$
valid for $N\rightarrow +\infty$, we obtain
\begin{eqnarray}
\label{exact12}
\frac{\partial \omega}{\partial t}+{\bf u}\cdot \nabla \omega=\nabla
\cdot (\nu\nabla \omega+\mu\gamma \omega\nabla \psi),
\end{eqnarray}
which is equivalent to Eq. (\ref{mf2}). In Fourier space, the mean field Fokker-Planck equation (\ref{exact12}) can be written
\begin{eqnarray}
\label{exact13}
\frac{\partial\hat{\omega}}{\partial t}({\bf k},t)+(2\pi)^{2}\int {\bf k}\cdot {\bf k}'_{\perp} \hat{\omega}({\bf k}-{\bf k}',t)\hat{u}({\bf k}')\hat{\omega}({\bf k}',t)d{\bf k}'\nonumber\\
=-\nu k^2 \hat{\omega}({\bf k},t)- (2\pi)^2\mu\gamma\int {\bf k}\cdot {\bf k}'\hat{\omega}({\bf k}-{\bf k}',t)\hat{u}({\bf k}')\hat{\omega}({\bf k}',t)d{\bf k}',
\end{eqnarray}
where we have noted ${\bf k}_{\perp}={\bf z}\times {\bf k}$. Let us
consider some particular cases of Eq. (\ref{exact12}). For
$\nu=\mu=0$, we find that the smooth vorticity field is solution of
the 2D Euler equation
\begin{eqnarray}
\label{exact14}
\frac{\partial\omega}{\partial t}+{\bf u}\cdot \nabla\omega=0.
\end{eqnarray}
This is the counterpart of the Vlasov equation in plasma physics. On
the other hand, for the model considered by Marchioro \&
Pulvirenti \cite{mp} where $\mu=0$ and
$\nu>0$, the smooth vorticity field is solution of the
Navier-Stokes equation
\begin{eqnarray}
\label{exact15}
\frac{\partial\omega}{\partial t}+{\bf u}\cdot
\nabla\omega=\nu\Delta\omega.
\end{eqnarray}
The derivation of this mean field equation was proven rigorously in \cite{mp}.

Equation (\ref{exact12}) for the ensemble averaged vorticity field
$\omega({\bf r},t)$ is a deterministic equation since we have averaged
over the noise. In contrast, Eq. (\ref{exact2}) for the exact
vorticity field $\omega_{d}({\bf r},t)$ is a stochastic equation
taking into account fluctuations. However, it is not very useful in
practice since the field $\omega_{d}({\bf r},t)$ is a sum of Dirac
distributions, not a regular function. Therefore, it is easier to
directly solve the stochastic equations (\ref{sto2}) rather than the
equivalent Eq. (\ref{exact2}). Following
\cite{ar,prep}, we can keep track of fluctuations while avoiding the
problem of $\delta$-functions by defining a ``coarse-grained''
vorticity field $\overline{\omega}({\bf r},t)$ obtained by averaging
the exact vorticity field on a spatio-temporal window of finite
resolution. For a weak long-range potential of interaction
and for a sufficiently small spatio-temporal window, we propose to
make the approximation
$\overline{\omega}^{(2)}({\bf r},{\bf r}',t)\simeq
\overline{\omega}({\bf r},t)\overline{\omega}({\bf r}',t)$. Then, we find
that the ``coarse-grained'' vorticity field satisfies a stochastic
equation of the form
\begin{eqnarray}
\label{exact16}
\frac{\partial\overline{\omega}}{\partial t}+\overline{\bf u}\cdot
\nabla\overline{\omega}=\nu\Delta\overline{\omega} +\nabla\cdot (\mu\gamma
\overline{\omega}\nabla\overline{\psi})+\nabla\cdot
(\sqrt{2\nu\gamma\overline{\omega}}\, {\bf R}).
\end{eqnarray}
If we ignore the noise term, Eq. (\ref{exact16}) reduces to
Eq. (\ref{mf2}). In that case, the system tends to a steady state that
is a maximum (global or local) of the free energy functional
$J[\overline{\omega}]$ at fixed circulation (minima or saddle points
of free energy are linearly dynamically unstable).  If the free energy
admits several local maxima (metastable states), the selection of the
steady state will depend on a notion of {\it basin of
attraction}. Without noise, the system remains on a maximum of free
energy forever. Now, in the presence of noise, the fluctuations can
induce {\it dynamical phase transitions} from one maximum to the
other. Thus, accounting correctly for fluctuations is very important
when there exists metastable states (see discussion in Sec. 2.3 of 
\cite{prep}).  This would be an interesting effect to study in more
detail.

\subsection{Newtonian or Coulombian potential}
\label{sec_newton}

Let us make the previous results  more explicit by considering
particular forms of potentials of interaction. The usual potential
of interaction between point vortices is solution of the Poisson
equation
\begin{eqnarray}
\label{newton1}
\Delta u=-\delta({\bf x}).
\end{eqnarray}
It corresponds to a Newtonian (or Coulombian) potential in two
dimensions. For box-confined vortices interacting through this
potential, it can be shown that the Gibbs canonical distribution
exists (i.e. the partition function is finite) if, and only, if
\cite{caglioti,kiessling,diff}
\begin{eqnarray}
\label{newton2}
\beta>\beta_c\equiv -\frac{8\pi}{N\gamma^2}.
\end{eqnarray}
At positive temperatures $\beta>0$, point vortices of the same sign
tend to ``repel'' each other and accumulate on the boundary of the
domain. This corresponds to a repulsive interaction like between
electric charges of the same sign in plasma physics. At negative
temperatures $\beta<0$, point vortices of the same sign tend to
``attract'' each other and form clusters. This corresponds to an
attractive interaction like between stars in a galaxy. As the inverse
temperature is reduced the cluster is more and more condensed and for
$\beta=\beta_c$, the equilibrium distribution is a Dirac peak
containing all the vortices. For $\beta<\beta_c$, there is no
equilibrium state (the Gibbs canonical distribution (\ref{sto1}) is not
normalizable and the free energy (\ref{ea4}) has no maximum). This regime can
be studied dynamically with the stochastic model (\ref{sto2})
introduced here.

In the mean field approximation valid for $N\rightarrow +\infty$, the
evolution of the smooth vorticity is described by the coupled system
\begin{eqnarray}
\label{newton3}
\frac{\partial \omega}{\partial t}+{\bf u}\cdot \nabla \omega=\nu
\nabla \cdot (\nabla \omega+\beta\gamma \omega\nabla \psi),
\end{eqnarray}
\begin{eqnarray}
\label{newton4}
-\Delta\psi=\omega.
\end{eqnarray}
The equilibrium state is obtained by solving the Boltzmann-Poisson
equation \cite{hou}.  By a proper reinterpretation of the
parameters, these equations are isomorphic to the 2D one-component
Smoluchowski-Poisson (SP) system, describing a gas of self-gravitating
Brownian particles
\cite{crs}, with an additional advective term. Like in gravity, we
expect that for $\beta\le \beta_c$, this system will describe the
collapse of the Brownian vortex gas. In fact, if we consider an
axisymmetric evolution, the advective term cancels out in
Eq. (\ref{newton3}) and we can immediately transpose the results
obtained in gravity for the SP system: (i) For $\beta>\beta_c$, the
system (\ref{newton3})-(\ref{newton4}) tends to an equilibrium state
in a bounded domain \cite{sc} and evaporates in an infinite domain
\cite{virial1}. (ii) For $\beta=\beta_c$, the system forms, in an
infinite time, a Dirac peak containing the $N$ point vortices. The
central vorticity diverges exponentially rapidly with time in a
bounded domain \cite{sc} and logarithmically in an unbounded domain
\cite{virial1}. (iii) For $\beta<\beta_c$, the system forms, in a
finite time $t_{coll}$, a Dirac peak containing $(\beta_c/\beta)N$
point vortices surrounded by a halo of vortices evolving pseudo
self-similarly \cite{sc}. A Dirac peak containing all the point
vortices is formed in the post-collapse regime $t>t_{coll}$ in a
finite time $t_{end}$ \cite{cmasse}.

Let us now determine the two-body correlation function of an infinite
and homogeneous distribution of point vortices using
Eq. (\ref{dh2}). In fact, for $\beta\neq 0$, a single species system
of point vortices with potential of interaction (\ref{newton1}) cannot
be spatially homogeneous. However, our approach can be justified by
the following arguments: (i) We can proceed {\it as if} the system
were infinite and homogeneous, making the equivalent of the ``Jeans
swindle'' in astrophysics \cite{jeans}. This can give us some hints
about the behaviour of the correlation functions for our system. (ii)
We can add a neutralizing background, like in the Jellium model of
plasma physics, so that spatially homogeneous equilibrium states of
point vortices now exist. (iii) We can consider a two-species system
and focus on one of the two species.  It will be shown in
Sec. \ref{sec_two} that the correlation functions of a two-components,
globally neutral, homogeneous system of point vortices are the {\it
same} as those derived here by making the ``Jeans swindle''. This
remark justifies a posteriori the validity of the following results
for a neutral system of point vortices. 

In an infinite domain, the
potential of interaction and its Fourier transform are given by
\begin{eqnarray}
\label{newton5}
u(x)=-\frac{1}{2\pi}\ln(x), \qquad (2\pi)^2
\hat{u}(k)=\frac{1}{k^2}.
\end{eqnarray}
Taking the divergence of Eq. (\ref{dh2}) and using the Poisson
equation (\ref{newton1}), we find that the two-body correlation
function satisfies an equation of the form
\begin{eqnarray}
\label{newton6}
\Delta h-\beta n\gamma^2 h=\beta \gamma^2 \delta({\bf x}).
\end{eqnarray}
The Fourier transform of the correlation function is
\begin{eqnarray}
\label{newton7}
n(2\pi)^2 \hat{h}(k)=\frac{-\beta n\gamma^2}{k^2+\beta n\gamma^2}.
\end{eqnarray}
According to the criterion (\ref{dh5}), the system is stable for
$\beta>0$ and unstable for $\beta<0$. At positive temperatures, a
point vortex of given circulation has the tendency to be surrounded by
point vortices of opposite circulation (in the two-species gas). This
is similar to Debye shielding in plasma physics and this allows the
existence of stable homogeneous states. At negative temperatures, an
infinite and homogeneous distribution of point vortices has the
tendency to collapse and form clusters. This is similar to the Jeans
instability in self-gravitating systems. Let us consider these two
cases consecutively.

At positive temperatures, it is convenient to introduce  the Debye
wavenumber $k_D=(\beta n \gamma^2)^{1/2}$. The two-body correlation
function can then be expressed as
\begin{eqnarray}
\label{newton8} 
n(2\pi)^2 \hat{h}(k)=\frac{-k_{D}^2}{k^2+k_{D}^2},\quad h(x)=-\frac{\beta
\gamma^2}{2\pi}K_{0}(k_{D}x).
\end{eqnarray}
This corresponds to the two-dimensional Debye-H\"uckel theory. The
correlation length is equal to the Debye length
\begin{eqnarray}
\label{newton9}
L_D=\frac{1}{k_{D}}=\frac{1}{(\beta n\gamma^2)^{1/2}}.
\end{eqnarray}
The energy of interaction (correlational energy) is given by 
\begin{eqnarray}
\label{newton10}
{E}=\frac{1}{2}n^2\gamma^2 V\int h({\bf x})u({\bf x})d{\bf x}, \quad {\rm or}\quad E=\frac{1}{2}n^2\gamma^2 V (2\pi)^2 \int \hat{h}({\bf k})\hat{u}({\bf k})d{\bf k}.
\end{eqnarray}
Evaluating the first integral in Eq.  (\ref{newton10}), which is
convergent, we get
\begin{eqnarray}
\label{newton12r}
E=-\frac{N\gamma^2}{8\pi}\left\lbrack \ln \left (\frac{\beta N\gamma^2}{4}\right )+2\gamma_{E}\right\rbrack, \quad \beta
=\frac{4}{N\gamma^2} e^{-\frac{8\pi E}{N\gamma^2}-2\gamma_{E}},
\end{eqnarray}
where $\gamma_{E}=0.577...$ is Euler's constant \footnote{The second
integral in Eq. (\ref{newton10}) is divergent. If we introduce a lower
cut-off $k_{min}\sim V^{-1/2}$ taking into account the large but
finite extent of the system, we get $E=-n^2 \gamma^4 \beta
V/(4\pi)\int_{k_{min} }^{+\infty}
(1/{k^2})({1}/({k^2+k_{D}^2}))kdk$. This leads to the following
relation $E=-{N\gamma^2}/({8\pi})\ln \left (1+{\beta
N\gamma^2/8\pi}\right )$ or $\beta={8\pi}/({N\gamma^2}) (e^{-8\pi
E/(N\gamma^2)}-1)$ between the average energy and the inverse
temperature. For convenience, we have taken $k_{max}=(8\pi/V)^{1/2}$
in order to recover the critical temperature
$\beta_{c}=-8\pi/(N\gamma^2)$ for $E\rightarrow +\infty$.  However, we
recall that the above results are only valid for $\beta>0$. }. On the
other hand, using Eq. (\ref{newton8}), the spatial correlations of the
vorticity in position space and its Fourier spectrum are given by (see
Appendix
\ref{sec_id}):
\begin{eqnarray}
\label{newton13}
\langle \delta\omega({\bf r})\delta\omega({\bf r}')\rangle =n\gamma^2\biggl\lbrace \delta({\bf r}-{\bf r}')-\frac{\beta n\gamma^{2}}{2\pi} K_{0}\left ( k_{D}|{\bf r}-{\bf r}'|\right )\biggr\rbrace,
\end{eqnarray}
\begin{eqnarray}
\label{newton14}
\langle \delta \hat{\omega}_{{\bf k}} \delta \hat{\omega}_{{\bf
k}'}\rangle=\frac{n\gamma^{2}}{(2\pi)^{2}}\frac{1}{1+\frac{n\beta
\gamma^{2}}{k^2}}\delta({\bf k}+{\bf k}').
\end{eqnarray}
The spectrum is depressed at small $k$ (i.e. long-wave fluctuations
are reduced) corresponding to Debye shielding in plasma physics. The
velocity distribution $W({\bf V})$ and the spatial correlations
$\langle {\bf V}(0)\cdot {\bf V}({\bf r})\rangle$ of the velocity
field created by the point vortices are discussed in
\cite{epjb}.

At negative temperatures, we can introduce the equivalent  the Jeans
wavenumber $k_J=(-\beta n \gamma^2)^{1/2}$. The two-body correlation
function can then be expressed as
\begin{eqnarray}
\label{newton15} 
n(2\pi)^2 \hat{h}(k)=\frac{k_{J}^2}{k^2-k_{J}^2},\quad h(x)=\frac{\beta\gamma^2}{4}Y_{0}(k_{J}x).
\end{eqnarray}
We note that $h(x)=-\beta\gamma^2 u(x)$ for $x\rightarrow 0$. The
oscillatory behaviour of the correlation function $h(x)$, or the fact
that the denominator of $\hat{h}(k)$ is negative for $k<k_{J}$, is
related to the instability (collapse) of a homogeneous distribution of
point vortices at negative temperatures (see criterion
(\ref{dh5})). The system is unstable to large wavelengths, above the
Jeans length $L_{J}\sim k_{J}^{-1}$. When the system becomes spatially
inhomogeneous, the dynamics of clustering can be studied with the mean
field equations (\ref{newton3})-(\ref{newton4}).

\subsection{Yukawa or Rossby potential}
\label{sec_yukawa}

We now assume that the potential of interaction is solution of an
equation of the form
\begin{eqnarray}
\label{yukawa1}
 \Delta
u-k_{0}^{2}u=-\delta({\bf x}),
\end{eqnarray}
where $k_0^{-1}$ is the characteristic range of the potential.  This
screened Poisson equation describes geophysical flows in the so-called
Quasi-Geostrophic (QG) approximation. In that context, $k_0$ is the
Rossby wavenumber. The screened Poisson equation (\ref{yukawa1}) also
arises in the Hasegawa-Mima model \cite{hm} which describes low-frequency
drift waves in magnetized plasmas. In an infinite domain, the
potential of interaction and its Fourier transform are given by
\begin{eqnarray}
\label{yukawa2}
u(x)=\frac{1}{2\pi}K_{0}(k_{0}x), \quad (2\pi)^2
\hat{u}(k)=\frac{1}{k^2+k_{0}^2}.
\end{eqnarray}
For a spatially homogeneous system, using Eqs. (\ref{dh3}) and
(\ref{yukawa2}), the Fourier transform of the correlation function is
\begin{eqnarray}
\label{yukawa3}
n(2\pi)^2 \hat{h}(k)=\frac{-\beta n\gamma^2}{k^2+k_{0}^2+\beta
n\gamma^2}.
\end{eqnarray}
According to the stability criterion  (\ref{dh5}), the system is stable iff
\begin{eqnarray}
\label{yukawa4}
\beta>-\frac{k_{0}^2}{n\gamma^2}\equiv \beta_0.
\end{eqnarray}
We note that, for a screened Coulombian potential, stable homogeneous
distributions exist at negative temperatures, above a critical
temperature $\beta_0$ depending on the characteristic range $k_0^{-1}$
of the potential. In terms of this critical temperature, the
correlation function (\ref{yukawa3}) can be rewritten
\begin{eqnarray}
\label{yukawa5}
(2\pi)^2 \hat{h}(k)=\frac{-\beta
\gamma^2}{k^2+k_{0}^2(1-\beta/\beta_0)},\quad {h}(x)=-\frac{\beta
\gamma^2}{2\pi}K_{0}\left\lbrack k_{0}(1-\beta/\beta_0)^{1/2}x\right
\rbrack.
\end{eqnarray}
The correlation length is
\begin{eqnarray}
\label{yukawa6}
L=\frac{1}{k_{0}}(1-\beta/\beta_0)^{-1/2}.
\end{eqnarray}

Let us first consider the case of positive temperatures. Introducing
the Debye wavenumber $k_D=(\beta n \gamma^2)^{1/2}$, the correlation
function and the correlation length can be written
\begin{eqnarray}
\label{yukawa7}
n(2\pi)^2 \hat{h}(k)=\frac{-k_{D}^2}{k^2+k_{0}^{2}+k_{D}^2}, \qquad
L=\frac{1}{\sqrt{k_{0}^{2}+k_{D}^{2}}}.
\end{eqnarray}
At positive temperatures, the correlation length is {\it smaller} than
the Rossby length: $L\le L_0$ where $L_0=1/k_0$. This corresponds to
the formation of a {\it screening cloud} of opposite sign vortices
around each vortex due to collective effects (like in plasma
physics). For $T\rightarrow 0^+$, i.e. $\beta\rightarrow +\infty$, the
screening length coincides with the Debye length: $L=L_D\rightarrow
0$. For $T\rightarrow +\infty$, i.e. $\beta\rightarrow 0^{+}$,
collective effects are negligible and we have $h({\bf
x})=-\beta\gamma^2 u({\bf x})$ \cite{hb1} so that the correlation
length coincides with the Rossby length: $L=L_0$. At negative
temperatures, introducing the Jeans wavenumber $k_J=(-\beta n
\gamma^2)^{1/2}$, the correlation function and the correlation length
can be written
\begin{eqnarray}
\label{yukawa8}
n(2\pi)^2 \hat{h}(k)=\frac{k_{J}^2}{k^2+k_{0}^{2}-k_{J}^2}, \qquad
  L=\frac{1}{\sqrt{k_{0}^{2}-k_{J}^{2}}}.
\end{eqnarray}
At negative temperatures, the correlation length is {\it larger} than
the Rossby length: $L\ge L_0$. This regime corresponds to the
formation of a {\it reinforcing cloud} due to the clustering of point
vortices of the same sign (like in gravity). For $T\rightarrow
-\infty$, i.e.  $\beta\rightarrow 0^{-}$, collective effects are
negligible so that the correlation length coincides with the Rossby
length: $L=L_0$.  For $T\rightarrow T_0$, i.e. $\beta\rightarrow
\beta_0$, the correlation length diverges. For $\beta<\beta_0$, the
system is unstable to wavenumbers
\begin{eqnarray}
\label{yukawa9}
k<k_{max}\equiv k_0 (\beta/\beta_0-1)^{1/2}.
\end{eqnarray}
For $\beta=\beta_0$, we have $k_{max}=0$ and for $\beta\rightarrow
-\infty$, we have $k_{max}\simeq k_J\rightarrow +\infty$. For a
screened potential, since $k_{max}^2=k_J^2-k_0^2$, the lengthscale
at which the system becomes unstable is larger than the Jeans
length.

For a screened potential, the energy of interaction (\ref{newton10})
is given by
\begin{eqnarray}
\label{yukawa10}
E=-\frac{n^2 \gamma^4 \beta
V}{4\pi}\int_{0}^{+\infty}\frac{1}{k^2+k_{0}^2}
\frac{1}{k^2+k_{0}^2(1-\beta/\beta_{0})}k \, dk,
\end{eqnarray}
yielding
\begin{eqnarray}
\label{yukawa11}
E=-\frac{N\gamma^2}{8\pi}\ln\left (1-\beta/\beta_{0}\right ),\qquad
\frac{\beta}{\beta_0}=1-e^{-\frac{8\pi E}{N\gamma^2}}.
\end{eqnarray}
On the other hand, using Eq. (\ref{yukawa5}), the spatial correlations
of the vorticity in position space and its Fourier spectrum are given
by (see Appendix
\ref{sec_id}):
\begin{eqnarray}
\label{yukawa12}
\langle \delta\omega({\bf r})\delta\omega({\bf r}')\rangle =n\gamma^2\biggl\lbrace \delta({\bf r}-{\bf r}')-\frac{\beta n\gamma^{2}}{2\pi} K_{0}\left\lbrack k_{0}(1-\beta/\beta_{0})^{1/2}|{\bf r}-{\bf r}'|\right\rbrack \biggr\rbrace,
\end{eqnarray}
\begin{eqnarray}
\label{yukawa13}
\langle \delta \hat{\omega}_{{\bf k}} \delta \hat{\omega}_{{\bf
k}'}\rangle=\frac{n\gamma^{2}}{(2\pi)^{2}}\frac{1}{1+\frac{n\beta
\gamma^{2}}{k^2+k_{0}^2}}\delta({\bf k}+{\bf k}').
\end{eqnarray}
Interestingly, these results can also be obtained directly from the
density of states (for large negative energies) \cite{et,pl} or from
the partition function \cite{prepa} by making the random phase
approximation.  The complementary approach presented here, based on
an equilibrium BBGKY-like hierarchy, is simpler because it avoids the use
of complex analysis. In the present context of Brownian vortices
(canonical ensemble), the control parameter is the temperature $\beta$
while for usual vortices (microcanonical ensemble)
\cite{et,pl}, the control parameter is the energy $E$. Let us interpret
the results (\ref{yukawa12})-(\ref{yukawa13}) in terms of the
temperature making a parallel with the discussion given by \cite{et} in terms
of the energy. When $T=\infty$ (i.e. $\beta=0$), the spectrum
(\ref{yukawa13}) is flat as for a random distribution of
vortices. When $\beta>0$, the spectrum is depressed at small $k$
(i.e. long-wave fluctuations are reduced). This corresponds to Debye
shielding in plasma physics in which each particle is surrounded by a
cloud of opposite sign particles. When $\beta<0$, the spectrum is
enhanced at small $k$ (i.e. long-wave fluctuations are
increased). This corresponds to a form of anti-shielding in which each
particle is surrounded by a cloud of similar particles. This is
similar to the Jeans clustering in astrophysics.  The second term in
the correlation function (\ref{yukawa12}) arises from the vortex cloud
which, on average, surrounds each particle. The radius of the cloud is
$L\sim k_0^{-1} (1-\beta/\beta_0)^{-1/2}$ and the effective number of
vortices contained in the cloud (obtained by integrating
(\ref{yukawa12})) is $N_{eff} \sim (1-\beta_0/\beta)^{-1}$. For
$\beta>0$, the cloud is of opposite circulation to the vortex and
shields its effect.  For $\beta<0$, the cloud is of same circulation
as the vortex and enhances its effect. The effective number of
vortices in the cloud increases as the inverse temperature decreases
and it diverges for $\beta=\beta_0$. This corresponds to the formation
of large ``clumps'' or ``clusters'' of vortices.  For $\beta<\beta_0$,
the statistically homogeneous distribution of vortices becomes
unstable and forms clusters. When the system becomes spatially
inhomogeneous, the dynamics of clustering can be studied with the mean
field equation (\ref{newton3}) coupled to the screened Poisson equation
$\Delta\psi-k_{0}^{2}\psi=-\omega$.

\subsection{The different regimes and the orbit-averaged-Fokker-Planck equation} \label{sec_regimes}

Using the previous kinetic theory, we can qualitatively discuss the
different regimes of the dynamics of 2D Brownian vortices depending on
the values of $N$, $\nu$ and $\mu=\beta\nu$ (a similar discussion
applies to self-gravitating Brownian particles \cite{virial2} and to
the BMF model \cite{bco}). For $\nu=\mu=0$ and $N\rightarrow +\infty$,
Eq. (\ref{mf2}) reduces to the 2D Euler equation. Starting from an
unstable initial condition, the 2D Euler equation can undergo a
process of violent relaxation and form a quasi-stationary state (QSS)
on a short dynamical timescale $\sim t_D$ \cite{miller,rs}. Then,
finite $N$ effects are expected to drive the ``collisional''
relaxation of the point vortex gas towards the microcanonical
Boltzmann distribution [with temperature $\beta(E)$] on a longer
timescale $t_{relax}(N)$ whose scaling with $N$ is not yet known (see
discussion in \cite{bbgky,cl}). For $\nu> 0$, $\mu\neq 0$ and $N\gg
1$, the mean field Fokker-Planck equation (\ref{mf2}) drives the
system towards the canonical Boltzmann distribution (\ref{mf4}) [with
bath temperature $\beta$] on a diffusive timescale $t_{diff}\sim
R^2/\nu$ independent on $N$. We can distinguish three cases: (i) When
$t_D\ll t_{relax}(N)\ll t_{diff}$ (corresponding to $N$ finite and
$\nu,\mu\rightarrow 0$), the system undergoes violent relaxation
towards an Euler QSS, then evolves towards a {\it microcanonical quasi
stationary state} \cite{bco} due to ``collisions'' (i.e. correlations
due to finite $N$ effects) and finally relaxes towards the canonical
equilibrium state. (ii) When $t_D\ll t_{diff}\ll t_{relax}(N)$
(corresponding to $N\rightarrow
\infty$ and relatively small viscosity), the system undergoes
violent relaxation towards an Euler QSS, then relaxes towards the canonical
equilibrium state without forming a microcanonical quasi stationary
state. (iii) When $t_{diff}\ll t_D\ll t_{relax}(N)$ (corresponding
to large viscosity $\nu,\mu\rightarrow +\infty$), the system
undergoes a diffusive relaxation towards the canonical equilibrium state
without forming a QSS.

Let us consider more specifically the second case. We assume
that $N\rightarrow +\infty$ so that ``collisional'' effects are completely
negligible. We also assume that $\nu\rightarrow 0$ and $\mu\rightarrow
0$ (while $\beta\sim 1$). To a first approximation, the kinetic
equation (\ref{mf2}) reduces to the 2D Euler equation. We assume that
the system has reached a QSS as a result of violent relaxation. This
is a stable stationary solution of the 2D Euler equation of the form
$\omega=f(\psi)$, which depends only on the stream function. It is
reached on a few dynamical times $t_D$. If $\nu$ is finite, the system
will evolve on a longer timescale due to the terms of drift and
diffusion present in the Fokker-Planck current. If $\nu$
and $\mu$ are sufficiently small, the solution
$\omega({\bf r},t)$ will remain close, at any time, to a steady
solution of the 2D Euler equation of the form $\omega=\omega(\psi,t)$
that slowly evolves in time due to the effects of diffusion and
drift. Noting that
\begin{equation}
\label{regime1}
\frac{\partial}{\partial t}\omega(\psi({\bf
r},t),t)=\frac{\partial\omega}{\partial
t}+\frac{\partial\psi}{\partial t}\frac{\partial\omega}{\partial
\psi},
\end{equation}
we can rewrite the kinetic equation  (\ref{mf2}) in the form
\begin{equation}
\label{regime2}
\frac{\partial\omega}{\partial t}+\frac{\partial\psi}{\partial
t}\frac{\partial\omega}{\partial \psi}=Q(\omega),
\end{equation}
where $Q(\omega)=-\nabla\cdot {\bf J}$ is the Fokker-Planck
term. Using the fact that $\omega$ only depends on the stream
function in a first approximation, we have
\begin{eqnarray}
\label{regime3}
Q(\omega)\equiv \nu \nabla \cdot (\nabla \omega+\beta\gamma
\omega\nabla \psi)=\nu\nabla\cdot\left \lbrack \left
(\frac{\partial\omega}{\partial \psi}+\beta\gamma\omega\right
)\nabla\psi\right\rbrack\nonumber\\
=\nu \left (\frac{\partial\omega}{\partial
\psi}+\beta\gamma\omega\right )\Delta\psi+\nu\nabla\psi\cdot \nabla
\left (\frac{\partial\omega}{\partial \psi}+\beta\gamma\omega\right
)\nonumber\\
=\nu \Delta\psi \left (\frac{\partial\omega}{\partial
\psi}+\beta\gamma\omega\right )+\nu (\nabla\psi)^2
\frac{\partial}{\partial\psi}\left (\frac{\partial\omega}{\partial
\psi}+\beta\gamma\omega\right ),
\end{eqnarray}
where $\Delta\psi=k_0^2\psi-\omega$ for the QG model
($\Delta\psi=-\omega$ for the ordinary point vortex model). The next
step is to get rid of the dependence on ${\bf r}$ in
Eq. (\ref{regime3}) by averaging over the orbits. Then, the dynamical
equation for $\omega(\psi,t)$ becomes
\begin{equation}
\label{regime4}
\frac{\partial\omega}{\partial t}+\left \langle
\frac{\partial\psi}{\partial t}\right\rangle\frac{\partial\omega}
{\partial \psi}=\nu (k_0^2\psi-\omega) \left
(\frac{\partial\omega}{\partial \psi}+\beta\gamma\omega\right )+\nu
\left \langle (\nabla\psi)^2\right\rangle
\frac{\partial}{\partial\psi}\left (\frac{\partial\omega}{\partial
\psi}+\beta\gamma\omega\right ),
\end{equation}
where $\langle X \rangle $ denotes the average of $X({\bf r},t)$ on
the elementary surface of the plane between $\psi$ and $\psi+d\psi$.
Defining
\begin{equation}
\label{regime5}
\chi(\psi,t)=\left \langle (\nabla\psi)^2\right\rangle=\frac{\int_{\psi\le \psi({\bf r},t)\le
\psi+d\psi}(\nabla\psi)^2d{\bf r}}{\int_{\psi\le \psi({\bf r},t)\le
\psi+d\psi}d{\bf r}}=\frac{\int \delta(\psi-\psi({\bf r},t))(\nabla\psi)^2 d{\bf r}}{\int \delta(\psi-\psi({\bf r},t)) d{\bf r}},
\end{equation}
and using  $\langle \partial\psi/\partial
t\rangle=\partial\langle\psi\rangle/\partial t=0$, we finally obtain
the orbit-averaged-Fokker-Planck equation
\begin{equation}
\label{regime6}
\frac{\partial}{\partial t}\omega(\psi,t)=\nu (k_0^2\psi-\omega)
\left (\frac{\partial\omega}{\partial \psi}+\beta\gamma\omega\right
)+\nu \chi(\psi,t) \frac{\partial}{\partial\psi}\left
(\frac{\partial\omega}{\partial \psi}+\beta\gamma\omega\right ),
\end{equation}
\begin{equation}
\label{regime7}
\Delta\psi=k_0^2\psi-\omega(\psi,t).
\end{equation}
The steady solution of this equation is the Boltzmann distribution 
(\ref{mf4}).

\section{The two-species system}
\label{sec_two}

\subsection{The stochastic equations} \label{sec_se}

For a multi-species system of point vortices, the stochastic equations consistent with
the Gibbs canonical distribution at statistical equilibrium are
\begin{equation}
\label{se1}
\frac{d{\bf r}_{i}}{dt}=-\frac{1}{\gamma_i}{\bf
z}\times\nabla_{i}H-\mu\nabla_iH+\sqrt{2\nu}{\bf R}_{i}(t),
\end{equation}
where $H=\sum_{i<j}\gamma_i \gamma_j u({\bf r}_{i},{\bf
r}_{j})+(1/2)\sum_{i=1}^N \gamma_i^2 v({\bf r}_i,{\bf r}_i)$ is the
Hamiltonian.  The Fokker-Planck
equation determining the evolution of the $N$-body distribution
function $P_N({\bf r}_1,...,{\bf r}_N,t)$ associated with the
stochastic equations (\ref{se1}) reads
\begin{eqnarray}
\label{se2}
\frac{\partial P_{N}}{\partial t}+\sum_{i=1}^{N}{\bf V}_{i}\cdot
\frac{\partial P_{N}}{\partial {\bf r}_{i}}
=\sum_{i=1}^{N}\frac{\partial}{\partial {\bf r}_{i}}\cdot\left
(\nu\frac{\partial P_{N}}{\partial {\bf r}_{i}}+\mu
P_{N}\frac{\partial H}{\partial {\bf r}_{i}}\right ),
\end{eqnarray}
where ${\bf V}_{i}=-(1/\gamma_i) {\bf z}\times \partial H/\partial
{\bf r}_{i}$ is the total advective velocity of point vortex $i$.
For $\nu\neq 0$ and $\mu\neq 0$, the stationary solution of the
Fokker-Planck equation (\ref{se2}) is the canonical distribution
\begin{eqnarray}
\label{se3}
P_{N}=\frac{1}{Z}e^{-\beta H},
\end{eqnarray}
provided that the mobility and the viscosity are related to each other
through the Einstein relation (\ref{ea3}). The free energy of the
$N$-body system is given by Eq. (\ref{ea4}) and we have the same
general properties as for the single species system (see
Sec. \ref{sec_ea}). If we restrict ourselves to {\it positive}
temperatures (i.e $\mu>0$), the stochastic equations (\ref{se1})
without the rotational term provide a microscopic description of the
Debye-H\"uckel model of electrolytes \cite{dh}. At positive
temperatures, the full equations (\ref{se1}) also provide a
microscopic description of a dissipative guiding center plasma under a
strong magnetic field, where like-sign charges repel each other (see
Appendix \ref{sec_gcp}). Here, we shall be more general and consider
both positive and negative temperatures, i.e repulsive or attractive
interactions between like-sign particles.

To be specific, let us consider a globally neutral two-species system
made of $N/2$ point vortices of circulation $+\gamma$ and $N/2$ point
vortices of circulation $-\gamma$. For box-confined vortices
interacting through the Newtonian (or Coulombian) potential
(\ref{newton1}), it can be shown that statistical equilibrium states
exist \footnote{The equation of state is known exactly for a 2D
point vortex gas (see, e.g.,
\cite{hou}). Note that the pressure must be positive at positive
temperatures $\beta>0$ and negative at negative temperatures
$\beta<0$. This determines a range of forbidden temperatures where the
pressure has not the right sign:
$\beta>\tilde{\beta}_{c}^{+}=8\pi/\gamma^2$ for neutral systems and
$\beta<\tilde{\beta}_{c}^{-}=-8\pi/\lbrack (N-1)\gamma^2\rbrack$ for a
single species system of $N$ point vortices. According to inequalities
(\ref{newton2}) and (\ref{se4}), the partition function diverges {\it
strictly before} entering this range of temperatures, so these
states are not accessible anyway (see, e.g.,
\cite{diff}).}  in the canonical ensemble (i.e. the partition function is
finite) iff (see \cite{km,diff} and references therein):
\begin{eqnarray}
\label{se4}
\beta_c^{-}\equiv -\frac{16\pi}{N\gamma^2} < \beta <
\beta_c^{+}=\frac{4\pi}{\gamma^2}.
\end{eqnarray}
At positive temperatures $\beta>0$ (implying $\mu>0$), the interaction
is ``repulsive'': the point vortices of the same sign have the
tendency to repel each other while the point vortices of opposite sign
have the tendency to attract each other. At large temperatures (small
$\beta$) a vortex of a given circulation is surrounded by a cloud of
vortices with opposite circulation which screen the interaction.  This
is like the Debye shielding in plasma physics. At small temperatures
(large $\beta$) the vortices have the tendency to form pairs or
microscopic dipoles $(+,-)$ similar to ``atoms'' $(+e,-e)$ in plasma
physics. In this regime, the flow consists in a spatially homogeneous
distribution of $N/2$ individual dipoles. At the critical temperature
$\beta_c^+$, the most probable state is a gas of $N/2$ {\it singular
microscopic dipoles} $(+,-)$ where the vortices of each pair have
fallen on each other. This leads to $N/2$ Dirac peaks made of two
vortices of opposite circulation. Thus, the value of the critical
temperature $\beta_c^+$ can be understood by considering the
statistical mechanics of only two vortices (one with positive
circulation $+\gamma$ and one with negative circulation $-\gamma$) and
determining at which temperature the partition function ceases to be
normalizable. The critical temperature $\beta_c^+$ can thus be
obtained from Eq. (\ref{newton2}) by taking $N=2$ and changing the
sign (since the ``attraction'' of like-sign vortices at $\beta<0$
corresponds to the ``attraction'' of opposite-sign vortices at
$\beta>0$). For $\beta>\beta_c^+$, there is no statistical equilibrium
state anymore and this regime can be studied dynamically with the
stochastic model (\ref{se1}). At negative temperatures $\beta<0$
(implying $\mu<0$), the interaction is ``attractive'': the point
vortices of the same sign have the tendency to attract each other
while the point vortices of opposite sign have the tendency to repel
each other. At sufficiently small temperatures (sufficiently negative
$\beta$), the vortices have the tendency to form two macroscopic
clusters: one cluster of $N/2$ positive vortices and one cluster of
$N/2$ negative vortices. In this regime, the flow consists in a
spatially inhomogeneous distribution of point vortices forming a
macroscopic dipole. At the critical temperature $\beta_c^-$, the most
probable state is a {\it singular macroscopic dipole}
$(\frac{N}{2}+,\frac{N}{2}-)$ where the vortices of the same species
have fallen on each other, creating two Dirac peaks made of $N/2$
vortices. The value of the critical temperature $\beta_c^-$ can be
understood by considering the statistical mechanics of a single
species system of $N/2$ vortices and determining at which temperature
the partition function ceases to be normalizable. The critical
temperature $\beta_c^-$ can thus be obtained from Eq. (\ref{newton2})
by making the substitution $N\rightarrow N/2$. For $\beta<\beta_c^-$
there is no statistical equilibrium state anymore and this regime can
be studied dynamically with the stochastic model (\ref{se1}).

\subsection{The BBGKY-like hierarchy} \label{sec_bbgky}

For sake of generality, we first consider a two-species system made of
$N_{+}=Nn_+$ point vortices of circulation $\gamma_+$ (labeled from $i=1$ to
$n_+N$) and $N_{-}=Nn_-$ point vortices of circulation $\gamma_-$ (labeled
from $i=n_+N+1$ to $N$). We use the same notations and presentation as
in the {\it equilibrium} study of Pointin \& Lundgren \cite{pl}. The
one-body distributions are defined by
\begin{eqnarray}
\label{b1} P_1^+({\bf r}_1,t)=\int P_N({\bf r}_1,...,{\bf
r}_N,t)d{\bf r}_2...d{\bf r}_N,
\end{eqnarray}
\begin{eqnarray}
\label{b2} P_1^-({\bf r}_N,t)=\int P_N({\bf r}_1,...,{\bf
r}_N,t)d{\bf r}_1...d{\bf r}_{N-1}.
\end{eqnarray}
The two-body distributions $P_2^{++}({\bf r}_1,{\bf r}_2,t)$,
$P_2^{+-}({\bf r}_1,{\bf r}_2,t)=P_2^{-+}({\bf r}_2,{\bf r}_1,t)$,
$P_2^{--}({\bf r}_1,{\bf r}_2,t)$, and the higher-body distributions
are defined similarly. From the Fokker-Planck equation (\ref{se2}), it is
easy to derive the complete BBGKY-like hierarchy of the two-species
point vortex gas. The general equations are
\begin{eqnarray}
\label{b3}  {\partial \over\partial t}P_{j}^{p^+,n^-}(1,...,p+n,t)=\sum_{i=1}^{p}\frac{\partial}{\partial {\bf r}_{i}}\biggl ( \nu \frac{\partial P_{j}^{p^+,n^-}}{\partial {\bf r}_{i}}+\mu P_{j}^{p^+,n^-}\sum_{k=1,k\neq i}^{p}\gamma_{+}^2 \frac{\partial' u_{ki}}{\partial {\bf r}_{i}}\nonumber\\
+\mu P_{j}^{p^+,n^-}\sum_{k=p+1}^{j}\gamma_{+}\gamma_{-}\frac{\partial' u_{ki}}{\partial {\bf r}_{i}}+\mu N \left (n_{+}-\frac{p}{N}\right )\gamma_{+}^{2}\int P_{j+1}^{(p+1)^+,n^{-}}\frac{\partial' u_{ki}}{\partial {\bf r}_{i}}d{\bf r}_{k}\nonumber\\
+\mu N\left (n_{-}-\frac{n}{N}\right )\gamma_{+}\gamma_{-}\int P_{j+1}^{p^+,(n+1)^-}\frac{\partial' u_{ki}}{\partial {\bf r}_{i}}d{\bf r}_{k}+\frac{1}{2}\mu P_{j}^{p^+,n^-}\gamma_{+}^2 \frac{\partial'}{\partial {\bf r}_{i}}v(i,i)\biggr )+\sum_{i=p+1}^{j}(+\leftrightarrow -),
\end{eqnarray}
where $j=p+n$.  The first equation of the BBGKY-like hierarchy for
$P_1^+$ is
\begin{eqnarray}
\label{b4}  {\partial P_{1}^+\over\partial t}=\nu
{\partial\over\partial {\bf r}_{1}}\biggl\lbrack {\partial
P_{1}^+\over\partial {\bf r}_{1}}+\beta N \left (n_+
-\frac{1}{N}\right )\gamma_+^{2}\int
 {\partial' {u}_{12}\over\partial {\bf r}_{1}}P_{2}^{++}(1,2)
d{\bf r}_{2}\nonumber\\
+\beta N n_-\gamma_+\gamma_-\int
 {\partial' {u}_{12}\over\partial {\bf r}_{1}}P_{2}^{+-}(1,2)
 d{\bf r}_{2}+\frac{1}{2}\beta P_1^+\gamma_+^2\frac{\partial'}
 {\partial {\bf r}_1}v(1,1)\biggr\rbrack.
\end{eqnarray}
The equation for $P_1^-$ is obtained by interchanging $+$ and $-$. The
second equations of the BBGKY-like hierarchy for $P_2^{++}$ and
$P_2^{+-}$ are
\begin{eqnarray}
\label{b5}{\partial P_{2}^{++}\over\partial t}=\nu
{\partial\over\partial {\bf r}_{1}}\biggl\lbrack {\partial
P_{2}^{++}\over\partial {\bf r}_{1}}+\beta \gamma_+^{2}
P_{2}^{++}(1,2){\partial' {u}_{12}\over\partial {\bf r}_{1}} +\beta
N \left (n_+-\frac{2}{N}\right) \gamma_+^{2} \int {\partial'
{u}_{13}\over\partial {\bf r}_{1}}P_{3}^{+++}(1,2,3) d{\bf r}_{3}\nonumber\\
+\beta N n_- \gamma_+\gamma_- \int {\partial' {u}_{13}\over\partial
{\bf r}_{1}}P_{3}^{++-}(1,2,3) d{\bf r}_{3}+\frac{1}{2}\beta
P_2^{++}\gamma_+^2\frac{\partial'}
 {\partial {\bf r}_1}v(1,1)
\biggr\rbrack+(1\leftrightarrow 2),\qquad
\end{eqnarray}
\begin{eqnarray}
\label{b6}{\partial P_{2}^{+-}\over\partial t}=\nu
{\partial\over\partial {\bf r}_{1}}\biggl\lbrack {\partial
P_{2}^{+-}\over\partial {\bf r}_{1}}+\beta \gamma_+\gamma_-
P_{2}^{+-}(1,2){\partial' {u}_{12}\over\partial {\bf r}_{1}} +\beta
N \left (n_+-\frac{1}{N}\right) \gamma_+^{2} \int {\partial'
{u}_{13}\over\partial {\bf r}_{1}}P_{3}^{+-+}(1,2,3) d{\bf r}_{3}\nonumber\\
+\beta N \left (n_- -\frac{1}{N}\right ) \gamma_+\gamma_- \int
{\partial' {u}_{13}\over\partial {\bf r}_{1}}P_{3}^{+--}(1,2,3)
d{\bf r}_{3}+\frac{1}{2}\beta P_2^{+-}\gamma_+^2\frac{\partial'}
 {\partial {\bf r}_1}v(1,1)
\biggr\rbrack+(1\leftrightarrow 2).\qquad
\end{eqnarray}
The equations for $P_2^{--}$ and $P_2^{-+}$ are obtained by
interchanging $+$ and $-$. Next, inserting the Mayer decompositions
\begin{eqnarray}
\label{b7}P_2^{++}(1,2)=P_1^+(1)P_1^+(2)+P_2^{'++}(1,2),
\end{eqnarray}
\begin{eqnarray}
\label{b8}P_2^{+-}(1,2)=P_1^+(1)P_1^-(2)+P_2^{'+-}(1,2),
\end{eqnarray}
in Eq. (\ref{b4}), we obtain the exact equation
\begin{eqnarray}
\label{b9}  {\partial P_{1}^+\over\partial t}=\nu
{\partial\over\partial {\bf r}_{1}}\biggl\lbrack {\partial
P_{1}^+\over\partial {\bf r}_{1}}+\beta N \left (n_+
-\frac{1}{N}\right )\gamma_+^{2}P_1^+(1)\int
 {\partial' {u}_{12}\over\partial {\bf r}_{1}}P_{1}^{+}(2)
d{\bf r}_{2}\nonumber\\
+\beta N \left (n_+ -\frac{1}{N}\right )\gamma_+^{2}\int
 {\partial' {u}_{12}\over\partial {\bf r}_{1}}P_{2}^{'++}(1,2)
d{\bf r}_{2}+\beta N n_-\gamma_+\gamma_- P_1^+(1)\int
 {\partial' {u}_{12}\over\partial {\bf r}_{1}}P_{1}^{-}(2)
 d{\bf r}_{2}\nonumber\\
 +\beta N n_-\gamma_+\gamma_-\int
 {\partial' {u}_{12}\over\partial {\bf r}_{1}}P_{2}^{'+-}(1,2)
 d{\bf r}_{2}+\frac{1}{2}\beta P_1^+\gamma_+^2\frac{\partial'}
 {\partial {\bf r}_1}v(1,1)\biggr\rbrack.
\end{eqnarray}
Inserting the Mayer decompositions
\begin{eqnarray}
\label{b10}P_3^{+++}(1,2,3)=P_1^+(1)P_1^+(2)P_1^+(3)
+P_1^+(1)P_2^{'++}(2,3)+P_1^+(2)P_2^{'++}(1,3)\nonumber\\
+P_1^+(3)P_2^{'++}(1,2)+P_3^{'+++}(1,2,3),
\end{eqnarray}
\begin{eqnarray}
\label{b11}P_3^{++-}(1,2,3)=P_1^+(1)P_1^+(2)P_1^-(3)
+P_1^+(1)P_2^{'+-}(2,3)+P_1^+(2)P_2^{'+-}(1,3)\nonumber\\
+P_1^-(3)P_2^{'++}(1,2)+P_3^{'++-}(1,2,3),
\end{eqnarray}
\begin{eqnarray}
\label{b12}P_3^{+-+}(1,2,3)=P_1^+(1)P_1^-(2)P_1^+(3)
+P_1^+(1)P_2^{'-+}(2,3)+P_1^-(2)P_2^{'++}(1,3)\nonumber\\
+P_1^+(3)P_2^{'+-}(1,2)+P_3^{'+-+}(1,2,3),
\end{eqnarray}
\begin{eqnarray}
\label{b13}P_3^{+--}(1,2,3)=P_1^+(1)P_1^-(2)P_1^-(3)
+P_1^+(1)P_2^{'--}(2,3)+P_1^-(2)P_2^{'+-}(1,3)\nonumber\\
+P_1^-(3)P_2^{'+-}(1,2)+P_3^{'+--}(1,2,3),
\end{eqnarray}
in Eqs. (\ref{b5})-(\ref{b6}), and using Eq. (\ref{b9}) to simplify some terms, we obtain the
exact equations
\begin{eqnarray}
\label{b14}{\partial P_{2}^{'++}\over\partial t}=\nu
{\partial\over\partial {\bf r}_{1}}\biggl\lbrack {\partial
P_{2}^{'++}\over\partial {\bf r}_{1}}+\beta
\gamma_+^{2}P_1^+(1)P_1^+(2){\partial' {u}_{12}\over\partial {\bf
r}_{1}}+\beta \gamma_+^{2} P_{2}^{'++}(1,2){\partial'
{u}_{12}\over\partial {\bf r}_{1}}\nonumber\\
-\beta \gamma_+^{2} P_1^+(1)P_1^+(2)\int {\partial'
{u}_{13}\over\partial {\bf r}_{1}}P_{1}^{+}(3) d{\bf r}_{3}
\nonumber\\
+\beta N \left (n_+-\frac{2}{N}\right) \gamma_+^{2} P_1^+(1)\int
{\partial' {u}_{13}\over\partial {\bf r}_{1}}P_{2}^{'++}(2,3) d{\bf
r}_{3}
 -\beta \gamma_+^{2} P_1^+(2)\int {\partial'
{u}_{13}\over\partial {\bf r}_{1}}P_{2}^{'++}(1,3) d{\bf
r}_{3}\nonumber\\
+\beta N \left (n_+-\frac{2}{N}\right) \gamma_+^{2}
P_2^{'++}(1,2)\int {\partial' {u}_{13}\over\partial {\bf
r}_{1}}P_{1}^{+}(3) d{\bf r}_{3} +\beta N \left
(n_+-\frac{2}{N}\right) \gamma_+^{2} \int {\partial'
{u}_{13}\over\partial {\bf r}_{1}}P_{3}^{'+++}(1,2,3) d{\bf
r}_{3}\nonumber\\
+\beta N n_- \gamma_+\gamma_- P_{2}^{'++}(1,2)\int {\partial'
{u}_{13}\over\partial {\bf r}_{1}} P_1^-(3) d{\bf r}_{3} +\beta N
n_- \gamma_+\gamma_- \int {\partial' {u}_{13}\over\partial {\bf
r}_{1}}P_{3}^{'++-}(1,2,3) d{\bf r}_{3}\nonumber\\
+\beta N n_-
\gamma_+\gamma_- P_1^+(1)\int {\partial' {u}_{13}\over\partial {\bf
r}_{1}}P_{2}^{'+-}(2,3) d{\bf r}_{3} +\frac{1}{2}\beta
P_2^{'++}(1,2)\gamma_+^2\frac{\partial'}
 {\partial {\bf r}_1}v(1,1)
\biggr\rbrack+(1\leftrightarrow 2),\nonumber\\
\end{eqnarray}
\begin{eqnarray}
\label{b15}{\partial P_{2}^{'+-}\over\partial t}=\nu
{\partial\over\partial {\bf r}_{1}}\biggl\lbrack {\partial
P_{2}^{'+-}\over\partial {\bf r}_{1}}+\beta \gamma_+\gamma_-
P_1^+(1)P_1^-(2){\partial' {u}_{12}\over\partial {\bf r}_{1}}+\beta
\gamma_+\gamma_- P_{2}^{'+-}(1,2){\partial'
{u}_{12}\over\partial {\bf r}_{1}}\nonumber\\
+\beta N\left (n_+-\frac{1}{N}\right )\gamma_+^{2} P_1^+(1)\int
{\partial' {u}_{13}\over\partial {\bf r}_{1}}P_{2}^{'-+}(2,3) d{\bf
r}_{3} \nonumber\\
+\beta N \left (n_+-\frac{1}{N}\right) \gamma_+^{2}
P_2^{'+-}(1,2)\int {\partial' {u}_{13}\over\partial {\bf
r}_{1}}P_{1}^{+}(3) d{\bf r}_{3} +\beta N \left
(n_+-\frac{1}{N}\right) \gamma_+^{2} \int {\partial'
{u}_{13}\over\partial {\bf r}_{1}}P_{3}^{'+-+}(1,2,3) d{\bf r}_{3}\nonumber\\
-\beta  \gamma_+\gamma_- P_1^{+}(1)P_1^-(2)\int {\partial'
{u}_{13}\over\partial {\bf r}_{1}}P_{1}^{-}(3) d{\bf r}_{3} +\beta N
\left (n_--\frac{1}{N}\right ) \gamma_+\gamma_- P_1^+(1)\int
{\partial' {u}_{13}\over\partial {\bf r}_{1}}P_{2}^{'--}(2,3) d{\bf
r}_{3}\nonumber\\
-\beta \gamma_+\gamma_- P_1^-(2)\int {\partial'
{u}_{13}\over\partial {\bf r}_{1}}P_{2}^{'+-}(1,3) d{\bf r}_{3}
+\beta N \left (n_- -\frac{1}{N}\right ) \gamma_+\gamma_-
P_{2}^{'+-}(1,2)\int {\partial' {u}_{13}\over\partial {\bf r}_{1}}
P_1^-(3) d{\bf r}_{3}\nonumber\\
+\beta N \left
(n_--\frac{1}{N}\right ) \gamma_+\gamma_- \int {\partial'
{u}_{13}\over\partial {\bf r}_{1}}P_{3}^{'+--}(1,2,3) d{\bf r}_{3}
+\frac{1}{2}\beta P_2^{'+-}(1,2)\gamma_+^2\frac{\partial'}
 {\partial {\bf r}_1}v(1,1)
\biggr\rbrack+(1\leftrightarrow 2).\qquad
\end{eqnarray}
To order $1/N$ in the proper thermodynamic limit defined in Sec. \ref{sec_ea},
the foregoing equations reduce to
\begin{eqnarray}
\label{b16}  {\partial P_{1}^+\over\partial t}=\nu
{\partial\over\partial {\bf r}_{1}}\biggl\lbrack {\partial
P_{1}^+\over\partial {\bf r}_{1}}+\beta N \left (n_+
-\frac{1}{N}\right )\gamma_+^{2}P_1^+(1)\int
 {\partial' {u}_{12}\over\partial {\bf r}_{1}}P_{1}^{+}(2)
d{\bf r}_{2}\nonumber\\
+\beta N n_+ \gamma_+^{2}\int
 {\partial' {u}_{12}\over\partial {\bf r}_{1}}P_{2}^{'++}(1,2)
d{\bf r}_{2}+\beta N n_-\gamma_+\gamma_- P_1^+(1)\int
 {\partial' {u}_{12}\over\partial {\bf r}_{1}}P_{1}^{-}(2)
 d{\bf r}_{2}\nonumber\\
 +\beta N n_-\gamma_+\gamma_-\int
 {\partial' {u}_{12}\over\partial {\bf r}_{1}}P_{2}^{'+-}(1,2)
 d{\bf r}_{2}+\frac{1}{2}\beta P_1^+\gamma_+^2\frac{\partial'}
 {\partial {\bf r}_1}v(1,1)\biggr\rbrack,
\end{eqnarray}
\begin{eqnarray}
\label{b17}{\partial P_{2}^{'++}\over\partial t}=\nu
{\partial\over\partial {\bf r}_{1}}\biggl\lbrack {\partial
P_{2}^{'++}\over\partial {\bf r}_{1}}+\beta
\gamma_+^{2}P_1^+(1)P_1^+(2){\partial' {u}_{12}\over\partial {\bf
r}_{1}} -\beta \gamma_+^{2} P_1^+(1)P_1^+(2)\int {\partial'
{u}_{13}\over\partial {\bf r}_{1}}P_{1}^{+}(3) d{\bf
r}_{3}\nonumber\\
 +\beta N n_+ \gamma_+^{2} P_1^+(1)\int {\partial'
{u}_{13}\over\partial {\bf r}_{1}}P_{2}^{'++}(2,3) d{\bf r}_{3}
+\beta N n_+ \gamma_+^{2} P_2^{'++}(1,2)\int {\partial'
{u}_{13}\over\partial {\bf r}_{1}}P_{1}^{+}(3) d{\bf r}_{3}\nonumber\\
+\beta N n_- \gamma_+\gamma_- P_{2}^{'++}(1,2)\int {\partial'
{u}_{13}\over\partial {\bf r}_{1}} P_1^-(3) d{\bf r}_{3} +\beta N
n_- \gamma_+\gamma_- P_1^+(1)\int {\partial' {u}_{13}\over\partial
{\bf r}_{1}}P_{2}^{'+-}(2,3) d{\bf r}_{3}
\biggr\rbrack+(1\leftrightarrow 2),\nonumber\\
\end{eqnarray}
\begin{eqnarray}
\label{b18}{\partial P_{2}^{'+-}\over\partial t}=\nu
{\partial\over\partial {\bf r}_{1}}\biggl\lbrack {\partial
P_{2}^{'+-}\over\partial {\bf r}_{1}}+\beta \gamma_+\gamma_-
P_1^+(1)P_1^-(2){\partial' {u}_{12}\over\partial {\bf r}_{1}} +\beta
N n_+\gamma_+^{2} P_1^+(1)\int {\partial' {u}_{13}\over\partial {\bf
r}_{1}}P_{2}^{'-+}(2,3) d{\bf r}_{3}\nonumber\\
 +\beta N
n_+\gamma_+^{2} P_2^{'+-}(1,2)\int {\partial' {u}_{13}\over\partial
{\bf r}_{1}}P_{1}^{+}(3) d{\bf r}_{3} -\beta  \gamma_+\gamma_-
P_1^{+}(1)P_1^-(2)\int {\partial'
{u}_{13}\over\partial {\bf r}_{1}}P_{1}^{-}(3) d{\bf r}_{3}\nonumber\\
+\beta N n_- \gamma_+\gamma_- P_1^+(1)\int {\partial'
{u}_{13}\over\partial {\bf r}_{1}}P_{2}^{'--}(2,3) d{\bf r}_{3}
+\beta N n_-  \gamma_+\gamma_- P_{2}^{'+-}(1,2)\int {\partial'
{u}_{13}\over\partial {\bf r}_{1}} P_1^-(3) d{\bf r}_{3}
\biggr\rbrack+(1\leftrightarrow 2).\nonumber\\
\end{eqnarray}
Before considering simplified forms of the above equations, we shall
derive useful expressions of the average energy.

\subsection{The average energy} \label{sec_ae}

The average energy
\begin{eqnarray}
\label{ae1} E=\langle H\rangle=\sum_{i<j}\int \gamma_i\gamma_j
u_{ij}P_N d{\bf r}_1...d{\bf r}_N+\frac{1}{2}\sum_{i=1}^N\int
\gamma_i^2 v_{ii}P_N d{\bf r}_1...d{\bf r}_N,
\end{eqnarray}
can be written
\begin{eqnarray}
\label{ae2} E=\frac{1}{2}\gamma_+^2 N_+ (N_+-1)\int P_2^{++}(1,2)
u_{12}d{\bf r}_1 d{\bf r}_2+ \frac{1}{2}\gamma_-^2 N_- (N_--1)\int
P_2^{--}(1,2) u_{12}d{\bf r}_1
d{\bf r}_2\nonumber\\
+N_+ N_- \gamma_+\gamma_- \int P_2^{+-}(1,2) u_{12}d{\bf r}_1 d{\bf
r}_2 +\frac{1}{2}\int \left\lbrack N_+ \gamma_+^2
P_1^+(1)+N_-\gamma_-^2P_1^-(1)\right\rbrack v(1,1) d{\bf r}_1.
\end{eqnarray}
Substituting the Mayer decomposition in the previous equation, we
obtain
\begin{eqnarray}
\label{ae3} E=\frac{1}{2}\gamma_+^2 N^2 n_+ \left
(n_+-\frac{1}{N}\right )\int \left\lbrack P_1^{+}(1)
P_1^{+}(2)+P_2^{'++}(1,2)\right\rbrack  u_{12}d{\bf r}_1 d{\bf
r}_2\nonumber\\
+\frac{1}{2}\gamma_-^2 N^2 n_- \left (n_--\frac{1}{N}\right )\int
 \left\lbrack P_1^{-}(1)
P_1^{-}(2)+P_2^{'--}(1,2)\right\rbrack u_{12}d{\bf r}_1 d{\bf
r}_2\nonumber\\
+N^2 n_+ n_-  \gamma_+\gamma_- \int \left\lbrack P_1^{+}(1)
P_1^{-}(2)+P_2^{'+-}(1,2)\right\rbrack u_{12}d{\bf r}_1 d{\bf r}_2
\nonumber\\
+\frac{N}{2}\int \left\lbrack n_+ \gamma_+^2
P_1^+(1)+n_-\gamma_-^2P_1^-(1)\right\rbrack v(1,1) d{\bf r}_1.
\end{eqnarray}
Introducing the average vorticity
\begin{eqnarray}
\label{ae4} \omega({\bf r},t)=N n_+\gamma_+ P_{1}^{+}({\bf r},t)
 +N n_-\gamma_- P_{1}^{-}({\bf r},t),
\end{eqnarray}
and the corresponding stream function (\ref{mf3}), the average energy can be
put in the form
\begin{eqnarray}
\label{ae5} E=\frac{1}{2}\int \omega\psi d{\bf
r}-\frac{N}{2}\int  \left\lbrack n_+ \gamma_+^2 P_2^{'++}(1,2)+n_-
\gamma_-^2 P_2^{'--}(1,2)\right\rbrack u_{12} d{\bf r}_1 d{\bf r}_2\nonumber\\
+\frac{1}{2} N^2\int \left\lbrack  \gamma_+^2 n_+^2
P_2^{'++}(1,2)+2 n_+ n_- \gamma_+ \gamma_- P_2^{'+-}(1,2)+
\gamma_-^2 n_-^2 P_2^{'--}(1,2)\right\rbrack u_{12}d{\bf r}_1 d{\bf
r}_2\nonumber\\
-\frac{1}{2}\gamma_+^2 N n_+ \int P_1^{+}(1) P_1^{+}(2)u_{12}d{\bf
r}_1 d{\bf r}_2-\frac{1}{2}\gamma_-^2 N n_- \int P_1^{-}(1)
P_1^{-}(2)u_{12}d{\bf r}_1 d{\bf r}_2 
\nonumber\\
+\frac{N}{2}\int \left\lbrack n_+ \gamma_+^2
P_1^+(1)+n_-\gamma_-^2P_1^-(1)\right\rbrack v(1,1) d{\bf r}_1.
\end{eqnarray}
This equation is exact for any $N$. At the order $1/N$ in the proper
thermodynamic limit defined in Sec. \ref{sec_ea}, it reduces to
\begin{eqnarray}
\label{ae6} E=\frac{1}{2}\int \omega\psi d{\bf
r}+\frac{N}{2}\int \left\lbrack n_+ \gamma_+^2
P_1^+(1)+n_-\gamma_-^2P_1^-(1)\right\rbrack v(1,1) d{\bf r}_1\nonumber\\
+\frac{1}{2} N^2\int \left\lbrack  \gamma_+^2 n_+^2 P_2^{'++}(1,2)+2
n_+ n_- \gamma_+ \gamma_- P_2^{'+-}(1,2)+ \gamma_-^2 n_-^2
P_2^{'--}(1,2)\right\rbrack u_{12}d{\bf r}_1 d{\bf
r}_2\nonumber\\
-\frac{1}{2}\gamma_+^2 N n_+ \int P_1^{+}(1) P_1^{+}(2)u_{12}d{\bf
r}_1 d{\bf r}_2-\frac{1}{2}\gamma_-^2 N n_- \int P_1^{-}(1)
P_1^{-}(2)u_{12}d{\bf r}_1 d{\bf r}_2.
\end{eqnarray}

\subsection{The mean field approximation} \label{sec_app}

In the limit $N\rightarrow +\infty$, the first equation (\ref{b16}) of the
BBGKY-like hierarchy reads
\begin{eqnarray}
\label{app1}  {\partial P_{1}^+\over\partial t}=\nu
{\partial\over\partial {\bf r}_{1}}\biggl\lbrace {\partial
P_{1}^+\over\partial {\bf r}_{1}}+\beta \gamma_+ P_1^+(1)\int
 {\partial' {u}_{12}\over\partial {\bf r}_{1}}\left\lbrack N n_+\gamma_+ P_{1}^{+}(2)
 +N n_-\gamma_- P_{1}^{-}(2)\right\rbrack
 d{\bf r}_{2}\biggr\rbrace.
\end{eqnarray}
Introducing the average vorticity (\ref{ae4}) and the corresponding stream
function (\ref{mf3}), we can rewrite Eq. (\ref{app1}) in the form
\begin{eqnarray}
\label{app2}  {\partial P_{1}^+\over\partial t}+{\bf u}\cdot\nabla
P_1^+=\nu \nabla\cdot \left (\nabla P_{1}^++\beta \gamma_+ P_1^+
\nabla\psi \right ),
\end{eqnarray}
where ${\bf u}=-{\bf z}\times \nabla\psi$ is the mean advective
velocity. The equation for $P_1^-$ is obtained by interchanging $+$
and $-$. Then, defining $\omega_+=N n_+\gamma_+ P_{1}^{+}$ and
$\omega_-=N n_-\gamma_- P_{1}^{-}$ so that $\omega=\omega_+
+\omega_-$, and considering a potential of interaction of the form
(\ref{newton1}), we obtain the coupled system of equations
\begin{eqnarray}
\label{app3}  {\partial \omega_+\over\partial t}+{\bf
u}\cdot\nabla\omega_+=\nu \nabla\cdot \left (\nabla \omega_+ +\beta
\gamma_+ \omega_+ \nabla\psi \right )\equiv -\nabla\cdot {\bf J}_+,
\end{eqnarray}
\begin{eqnarray}
\label{app4}  {\partial \omega_-\over\partial t}+{\bf
u}\cdot\nabla\omega_-=\nu \nabla\cdot \left (\nabla \omega_- +\beta
\gamma_- \omega_- \nabla\psi \right )\equiv -\nabla\cdot {\bf J}_-,
\end{eqnarray}
\begin{eqnarray}
\label{app5}  -\Delta\psi=\omega_+ +\omega_-,
\end{eqnarray}
with the normalization $\int \omega_+ d{\bf r}=N n_+ \gamma_+=\Gamma_{+}$ and
$\int \omega_- d{\bf r}=N n_- \gamma_-=\Gamma_{-}$. These mean field
Fokker-Planck equations conserve the circulations (or vortex numbers) of
each species. The steady solutions are the Boltzmann distributions
\begin{eqnarray}
\label{app6}  \omega_+=A_+ e^{-\beta \gamma_+ \psi},\qquad
\omega_-=A_- e^{-\beta \gamma_- \psi}.
\end{eqnarray}
They can also be obtained from the equilibrium BBGKY hierarchy in the
limit $N\rightarrow +\infty$. Substituting Eq. (\ref{app6}) in
Eq. (\ref{app5}), we obtain a mean field equation determining the
equilibrium stream function. For a globally neutral system with
$\gamma_{+}=-\gamma_{-}=\gamma$ and $n_{+}=n_{-}=1/2$, this equation
reduces, in the symmetric case $A_+=-A_-$, to the celebrated
sinh-Poisson equation $\Delta\psi=2A\sinh(\beta\gamma\psi)$ studied in
\cite{jm,tcl} (note, however, that there is no fundamental reason why
$A_{+}=-A_{-}$; this corresponds to very particular initial
conditions). At positive temperatures, and for a globally neutral
system, the kinetic equations (\ref{app3})-(\ref{app5}), without the
advective term, are isomorphic to the Debye-H\"uckel model of
electrolytes where like-sign charges repel each other (see Eq. (3) of
\cite{dh}). Therefore, for $\beta>0$, our study provides a kinetic
derivation, from the microscopic stochastic process (\ref{se1}), of
the mean field kinetic equations introduced by Debye \& H\"uckel
\cite{dh}. At positive temperatures, these equations lead at
statistical equilibrium to a spatially uniform state with zero net
charge. The novelty of our model is to allow also for the
consideration of negative temperatures where like-sign vortices
attract each other. For sufficiently negative inverse temperatures,
these equations lead to a macroscopic order, typically the formation
of a large-scale dipole \cite{rc}.

The Boltzmann entropy of the two-species point vortex gas
\begin{eqnarray}
\label{app7}  S=-N\int \left\lbrack n_+ P_1^+(1)\ln P_1^+(1)+n_-
P_1^-(1)\ln P_1^-(1)\right\rbrack d{\bf r}_1,
\end{eqnarray}
can be rewritten (up to an additive constant):
\begin{eqnarray}
\label{app8}  S=-\int \left (\frac{\omega_+}{\gamma_+}
\ln\frac{\omega_+}{\gamma_+} + \frac{\omega_-}{\gamma_-}
\ln\frac{\omega_-}{\gamma_-}\right ) d{\bf r}.
\end{eqnarray}
This expression of the Boltzmann entropy can also be obtained from a
classical combinatorial analysis \cite{cl}. On the other hand, for $N\rightarrow +\infty$, the mean field energy
is
\begin{eqnarray}
\label{app9}  E=\frac{1}{2}\int \omega \psi d{\bf r}.
\end{eqnarray}
Then,  the mean field free energy (more properly the Massieu
function) reads
\begin{eqnarray}
\label{app10}  J=S-\beta E,
\end{eqnarray}
where $S$ and $E$ are given by Eqs. (\ref{app8}) and (\ref{app9}). The
mean field free energy (\ref{app10}) can be obtained from
Eq. (\ref{ea4}) in the limit $N\rightarrow +\infty$. It is
straightforward to establish that
\begin{eqnarray}
\label{app11}  \dot J=\int \left (\frac{{\bf J}_+^2}{\nu \gamma_+
\omega_+}+\frac{{\bf J}_-^2}{\nu \gamma_- \omega_-}\right )d{\bf r}\ge 0.
\end{eqnarray}
This is the appropriate form of $H$-theorem in the canonical ensemble:
$\dot J\ge 0$ and $\dot J=0$ iff $\omega_{\pm}$ are the mean field
Boltzmann distributions (\ref{app6}). The free energy
$J[\omega_{+},\omega_{-}]$ is the Lyapunov functional of the
Fokker-Planck equations (\ref{app3})-(\ref{app5}). The Boltzmann
distributions (\ref{app6}) are critical points of $J$ at fixed
$\Gamma_{+}$ and $\Gamma_{-}$ and they are linearly dynamically stable
iff they are (local) maxima of $J$. If $J$ is bounded from above,
we conclude from Lyapunov's direct method, that the mean field
Fokker-Planck equations (\ref{app3})-(\ref{app5}) will reach, for
$t\rightarrow +\infty$, a (local) maximum of $J$ at fixed circulations
$\Gamma_+$ and $\Gamma_-$. If several local maxima exist (like dipoles
and bars in a periodic domain
\cite{ymc}), the selection of the maximum will depend on a complicated
notion of basin of attraction. If there is no maximum of free energy
at fixed circulations, the system will have a peculiar behavior and
will generate singularities. This is the case in particular for
$\beta<\beta_c^-$ in the situation described in Sec. \ref{sec_se}. In
that case, the partition function diverges and the mean field free
energy has no global maximum. This situation can be studied
dynamically by solving the mean field Fokker-Planck equations
(\ref{app3})-(\ref{app5}). Preliminary numerical simulations \cite{rc}
of Eqs. (\ref{app3})-(\ref{app5}) show the formation of a {\it
singular macroscopic dipole} \footnote{Note that a similar growing
condensate (dipole) has been studied recently in the context of forced
2D turbulence \cite{lebedev}. Our model of Brownian vortices is
physically different from \cite{lebedev} but the connexion between the
two studies may be noted.}. The results of these simulations will be
reported elsewhere \cite{rc}.  We note that the relaxation equations
(\ref{app3})-(\ref{app5}) can be obtained from a maximum entropy
production principle (MEPP) by maximizing the production of free
energy $\dot J$ at fixed circulations $\Gamma_{\pm}$ (and additional
physical constraints) \cite{prepa}.

It is also straightforward to generalize
the present results to a multi-species system of Brownian vortices. The general
mean field Fokker-Planck equations are given by 
\begin{eqnarray}
\label{app12}
\frac{\partial \omega_{\alpha}}{\partial t}+{\bf
u}\cdot\nabla\omega_{\alpha}=\nu \nabla\cdot
(\nabla\omega_{\alpha}+\beta\gamma_{\alpha}\omega_{\alpha}\nabla\psi),
\end{eqnarray}
where $\psi$ is given by Eq. (\ref{mf3}) with
$\omega=\sum_{\alpha}\omega_{\alpha}$ where $\omega_{\alpha}({\bf
r},t)=N_{\alpha}\gamma_{\alpha}P_{1}^{(\alpha)}({\bf r},t)$ is the
ensemble averaged vorticity (proportional to the density) of species
$\alpha$.  The free energy (Massieu function) is $J=S-\beta E$ where
$S=-\sum_{\alpha}\int
\frac{\omega_{\alpha}}{\gamma_{\alpha}}\ln\frac{\omega_{\alpha}}{\gamma_{\alpha}}\,
d{\bf r}$ and $E$ is given by Eq. (\ref{app9}). The $H$-theorem reads
\begin{equation}
\dot J=\nu\sum_{\alpha}\int \frac{1}{\gamma_{\alpha}\omega_{\alpha}}(\nabla\omega_{\alpha}+\beta\gamma_{\alpha}\omega_{\alpha}\nabla\psi)^2\, d{\bf r}.
\label{fdjdor}
\end{equation}
The relaxation equations (\ref{app12}) have the following properties:
(i) $\Gamma_{\alpha}=\int\omega_{\alpha}\, d{\bf r}$ are conserved
(ii) $\dot J\ge 0$ (iii) $\dot J=0$ iff
$\omega_{\alpha}=A_{\alpha}e^{-\beta\gamma_{\alpha}\psi}$ (iv) a
steady state is linearly stable iff it is a (local) maximum of $J$ at
fixed $\Gamma_{\alpha}$. Thus, if $J$ is bounded from above, these
equations will relax towards the canonical statistical equilibrium
state of the multi-species point vortex gas. We can also extend the
approach of Sec. \ref{sec_exact} to obtain the equations satisfied by
the exact vorticity field of each species. They are given by
\begin{eqnarray}
\label{app13}
\frac{\partial \omega_{\alpha}}{\partial t}+{\bf
u}\cdot\nabla\omega_{\alpha}=\nu\nabla\cdot
(\nabla\omega_{\alpha}+\beta\gamma_{\alpha}\omega_{\alpha}\nabla\psi)+\nabla\cdot (\sqrt{2\nu\gamma_{\alpha}\omega_{\alpha}}\ {\bf R}),
\end{eqnarray}
where $\psi$ is given by Eq. (\ref{exact3}) with
$\omega=\sum_{\alpha}\omega_{\alpha}$ where $\omega_{\alpha}({\bf
r},t)=\gamma_{\alpha}\sum_{i\in X_{\alpha}}\delta({\bf r}-{\bf
r}_{i}(t))$ is the exact vorticity field of species $\alpha$ (here,
$X_{\alpha}$ denotes the ensemble of point vortices of species
$\alpha$).
Using the same arguments as those given at the end of
Sec. \ref{sec_exact}, we can introduce a coarse-grained vorticity
field that smoothes out the Dirac distributions while keeping track of
stochastic effects. The evolution of the coarse-grained vorticity is
then given by an equation of the form (\ref{app13}) where $\psi$ is
given by Eq. (\ref{mf3}) with $\omega=\sum_{\alpha}\omega_{\alpha}$
where $\omega_{\alpha}({\bf r},t)$ denotes now the coarse-grained
vorticity of species $\alpha$. These stochastic kinetic equations,
involving a noise term, could be used to study dynamical phase
transitions between different maxima of free energy, like ``bars'' and
``dipoles'' in periodic domains. This will be considered in future
contributions.

\subsection{The two-body correlation function} \label{sec_corr}

We now consider a globally neutral system such that
$\gamma_+=+\gamma$, $\gamma_-=-\gamma$ and $n_+=n_-=1/2$. We also
assume that the distribution of point vortices is spatially
homogeneous (to leading order) so that $P_1^\pm({\bf
r},t)=P_0+\hat{P}^{\pm}({\bf r},t)$ where $\hat{P}^{\pm}$ is of order
$1/N$. In that case, the second equations (\ref{b17})-(\ref{b18}) of
the BBGKY hierarchy at the order $1/N$ reduce to the forms
\begin{eqnarray}
\label{c1}{\partial P_{2}^{'++}\over\partial t}=\nu
{\partial\over\partial {\bf r}_{1}}\biggl\lbrace {\partial
P_{2}^{'++}\over\partial {\bf r}_{1}}+\beta \gamma^{2}P_0^2
{\partial' {u}_{12}\over\partial {\bf r}_{1}} -\beta \gamma^{2}
P_0^3\int {\partial' {u}_{13}\over\partial {\bf r}_{1}} d{\bf
r}_{3}\nonumber\\
 +\frac{1}{2}\beta N \gamma^{2} P_0\int {\partial'
{u}_{13}\over\partial {\bf r}_{1}}\left\lbrack
P_{2}^{'++}(2,3)-P_2^{'+-}(2,3)\right\rbrack  d{\bf
r}_{3}\biggr\rbrace +(1\leftrightarrow 2),
\end{eqnarray}
\begin{eqnarray}
\label{c2}{\partial P_{2}^{'+-}\over\partial t}=\nu
{\partial\over\partial {\bf r}_{1}}\biggl\lbrace {\partial
P_{2}^{'+-}\over\partial {\bf r}_{1}}-\beta \gamma^{2}P_0^2
{\partial' {u}_{12}\over\partial {\bf r}_{1}} +\beta \gamma^{2}
P_0^3\int {\partial' {u}_{13}\over\partial {\bf r}_{1}} d{\bf
r}_{3}\nonumber\\
 +\frac{1}{2}\beta N \gamma^{2} P_0\int {\partial'
{u}_{13}\over\partial {\bf r}_{1}}\left\lbrack
P_{2}^{'-+}(2,3)-P_2^{'--}(2,3)\right\rbrack  d{\bf
r}_{3}\biggr\rbrace +(1\leftrightarrow 2),
\end{eqnarray}
and the energy of interaction is
\begin{eqnarray} \label{c3} E=\frac{N^2\gamma^2}{8}\int \left\lbrack
P_2^{'++}(1,2)-2 P_2^{'+-}(1,2)+ P_2^{'--}(1,2)\right\rbrack
u_{12}d{\bf r}_1 d{\bf r}_2 \nonumber\\
-\frac{N\gamma^2}{2} \int
P_0^2 u_{12}d{\bf r}_1 d{\bf r}_2 +\frac{1}{2}N\gamma^2 P_0\int
v(1,1) d{\bf r}_1.
\end{eqnarray}
Considering solutions such that (note the presence of the term $1/N$ in the definition of $h$): 
\begin{eqnarray}
\label{c4}
P_2^{'++}(1,2)=P_2^{'--}(1,2)=-P_2^{'+-}(1,2)=-P_2^{'-+}(1,2)=P_0^2
\left\lbrack h({\bf r}_1-{\bf r}_2,t)+\frac{1}{N}\right\rbrack,
\end{eqnarray}
we find that the two-body correlation function $h({\bf r}_1-{\bf
r}_2,t)$ satisfies an equation of the form
\begin{eqnarray}
\label{c5}{\partial h\over\partial t}=2\nu {\partial\over\partial
{\bf r}_{1}}\cdot \biggl\lbrace {\partial h\over\partial {\bf
r}_{1}}+\beta \gamma^{2} {\partial {u}_{12}\over\partial {\bf
r}_{1}}
 +\beta N \gamma^{2} P_0\int {\partial
{u}_{13}\over\partial {\bf r}_{1}}h(2,3) d{\bf r}_{3}\biggr\rbrace.
\end{eqnarray}
This returns Eq. (\ref{dh1}) for a one
component system with a neutralizing background. The correlational
energy is
\begin{eqnarray} \label{c6} \tilde{E}\equiv E-\frac{1}{2}N\gamma^2 P_0\int
v(1,1) d{\bf r}_1= \frac{N^2\gamma^2}{2}P_0^2 \int h(1,2)
 u_{12}d{\bf r}_1 d{\bf r}_2,
\end{eqnarray}
which coincides with Eq. (\ref{newton10}). Finally, the first equation of the BBGKY-like hierarchy, written at
the order $1/N$, gives the evolution of the function $\hat{P}^{+}({\bf r},t)$
in the form
\begin{eqnarray}
\label{c7}  {\partial \hat{P}^+\over\partial t}=\nu
{\partial\over\partial {\bf r}_{1}}\biggl\lbrack {\partial
\hat{P}^+\over\partial {\bf r}_{1}}+\frac{1}{2}\beta
N\gamma^{2}P_0\int
 {\partial' {u}_{12}\over\partial {\bf r}_{1}}\left\lbrack
 \hat{P}^{+}(2)-\hat{P}^{-}(2)+2h(1,2)P_0\right\rbrack
d{\bf r}_{2}\nonumber\\
 -\beta \gamma^2P_0^2\int
 {\partial' {u}_{12}\over\partial {\bf r}_{1}}
 d{\bf r}_{2}+\frac{1}{2}\beta P_0\gamma^2\frac{\partial'}
 {\partial {\bf r}_1}v(1,1)\biggr\rbrack,
\end{eqnarray}
and an equivalent equation for $\hat{P}^-({\bf r},t)$. Since we
recover the same results as in Sec. \ref{sec_dh}, this two-species model
justifies, on a firmer basis, the results obtained in
Secs. \ref{sec_newton} and \ref{sec_yukawa} based on a ``Jeans-like
swindle''.  We note that the correlation function (\ref{newton8}) and
the caloric curve (\ref{newton12r}) are defined for {\it all} positive
temperatures. In particular, these formulae seem to be valid above the
critical inverse temperature $\beta_{c}^{+}$. This is of course
incorrect and this clearly shows that the Debye-H\"uckel theory is not
valid for low temperatures (note that the random phase approximation
in \cite{et} also breaks down at low temperatures). In particular, it
cannot account for the formation of tightly bound dipoles or pairs
$(+,-)$ as $\beta\rightarrow \beta_{c}^{+}$. To arrive at the
expressions (\ref{newton8}) and (\ref{newton12r}), we have assumed
that the three-body correlation function $P_{3}'$ could be neglected
because it is of order $1/N^2$. This is correct ``on average'' but
this is not correct {\it at any scale}, in particular for small
separations. Therefore, it is necessary to take into account higher
order correlation functions in the kinetic theory (and go beyond the
Debye-H\"uckel approximation) if we want to reproduce the pair
formation at low temperatures. In this respect, we note the important
paper of Samaj \& Travenec \cite{st} who obtained the exact
thermodynamic parameters of the 2D two-components plasma for any value
of accessible positive temperatures $0\le \beta\le
\beta_{c}^{+}$. This study clearly shows that the Debye-H\"uckel
theory is only valid at large temperatures.

\section{Conclusion}
\label{sec_conclusion}

In this paper (see also
\cite{crrs}) we have introduced a new model of random walkers in
long-range interactions that we called ``Brownian vortices''. By
definition, their dynamics is governed by stochastic equations of the
form (\ref{se1}). This dynamical model is associated with the
canonical ensemble and it leads to the Gibbs canonical distribution at
statistical equilibrium.  In the case of Brownian vortices, we can
develop a kinetic theory and analyze all stages of the dynamics.  In
the mean-field limit, we find that the smooth vorticity field is
governed by a parabolic drift-diffusion equation (\ref{mf2}) coupled
to an equation of the form (\ref{mf3}), which is often an elliptic
equation like the Poisson equation. This type of equations is actively
studied at present by physicists and applied mathematicians in
relation to self-gravitating Brownian particles and biological
populations experiencing chemotaxis \cite{banach}. We think that the
present model naturally enters in this line of investigations. In
particular, an interesting novelty of this model is to allow for
particles (vortices) with positive and negative circulations (while
the mass of material particles -like stars or bacteria- is always
positive) leading to a wider richness of phenomena.

Another interest of the point vortex model is to allow for both
positive and negative temperatures (see Table \ref{ana}). At negative
temperatures, the interaction between like-sign vortices is
``attractive'' (like gravity) and can create a macroscopic order
characterized by the nonuniformity of the spatial distribution of the
vortices (typically a monopole or a dipole). The statistical mechanics
of point vortices at negative temperatures shares a lot of analogies
with the statistical mechanics of stellar systems in astrophysics
\cite{hou}. It is, furthermore, richer because we can have positive
and negative circulations resulting in the separation of a cluster of
positive circulations and a cluster of negative circulation (dipole)
while gravitational systems typically organize in a single cluster. We
have also mentioned the existence of a critical inverse temperature
$\beta_c^{-}$ below which the partition function diverges and the free
energy is not bounded anymore. In that case, we expect the system of
point vortices to collapse. It forms a singular monopole $N\gamma$
(Dirac peak) in the single-species case and a singular dipole
$(\frac{N}{2}\gamma,-\frac{N}{2}\gamma)$ in the two-species case.  At
positive temperatures, the interaction between like-sign vortices is
``repulsive'' (like in electrostatics). For a neutral system, this
leads to a spatially homogeneous state with correlations which behaves
therefore like a $2D$ Coulombian plasma. We have mentioned the
existence of a critical inverse temperature $\beta_c^{+}$ above which
the partition function diverges and the free energy is not bounded
anymore. In that case, we expect the system of point vortices to
collapse. It forms a gas of $N/2$ singular dipoles $(+\gamma,-\gamma)$
similar to atoms $(+e,-e)$ in a 2D plasma.

For inhomogeneous systems, the mean field energy (\ref{app9}) scales
like $N^2$, i.e. the normalized energy $\epsilon=E/(N^2\gamma^2)$ is
of order unity, corresponding to a nonextensive scaling. This regime
of negative temperatures, leading to a macroscopic order (like for
self-gravitating systems), has been studied by Onsager \cite{onsager}
and others \cite{jm,mj,et,kida,pl,lp,caglioti,es,kl} at statistical
equilibrium.  On the other hand, at positive temperatures, the
interaction between like-sign vortices is ``repulsive'' (like for
electric charges) and the physics is very similar to that of a
two-dimensional plasma. In that case, we recover the model of
Debye-H\"uckel electrolytes \cite{dh}. In the neutral case, the
distribution of vortices is spatially homogeneous due to Debye
shielding. For homogeneous systems, the correlational energy
(\ref{c6}) scales like $N$ (since $P_2'\sim 1/N$), so it has an
extensive scaling. This regime of positive temperatures, where the
usual thermodynamic limit applies, has been studied by Ruelle
\cite{ruelle} and others \cite{et,fr} at statistical equilibrium. 
It was believed for some time that these two approaches (Onsager and
Ruelle) were in contradiction. In fact, they correspond to different
definitions of the thermodynamic limit describing different
situations. The present model of Brownian vortices provides an
out-of-equilibrium model that is fully consistent with the results of
statistical equilibrium in the canonical ensemble at both positive and
negative temperatures in the two situations described above.

The perspectives of this work are the following. On a
theoretical point of view, it would be important to study numerically,
analytically and with the methods of applied mathematics the mean
field Fokker-Planck equations (\ref{app3})-(\ref{app5}). This has
largely been done already in the single species case since the
corresponding equations (\ref{newton3})-(\ref{newton4}) are isomorphic
to the Smoluchowski-Poisson system and Keller-Segel model
\cite{crrs}. However, the two species mean field Fokker-Planck
equations (\ref{app3})-(\ref{app5}) have not been studied extensively
so far (especially in the attractive case valid in the regime of
negative temperatures). In particular, the formation of the singular
macroscopic dipole for $\beta<\beta_{c}^{-}$ is of particular
theoretical interest \cite{rc}. The study of the $N$-body stochastic
equations (\ref{sto2}) or (\ref{se1}) and the study of the stochastic
kinetic equation (\ref{exact16}) or (\ref{app13}) are also of
interest. When the free energy $J$ has several local maxima at fixed
circulation (metastable states), representing for example monopoles,
dipoles, bars or jets like in
\cite{ymc}, the stochastic term (noise)
in Eq. (\ref{exact16}) or (\ref{app13}) can induce {\it dynamical
phase transitions} from one state to the other.  It would be
interesting to study this phenomenon in detail. On the other hand, on
a physical point of view, it would be interesting to know if we can
experimentally realize a system of Brownian vortices described by
Eqs. (\ref{se1}).  It would be interesting to devise experiments (not
necessarily in the context of fluid mechanics) where the effective
coupling of the particles with a thermal bath can be modeled by
stochastic processes of the form (\ref{se1}). In the repulsive case,
these equations can describe a dissipative plasma (see
\cite{dh} and Appendix \ref{sec_gcp}). In the attractive case,
these equations could describe biological systems where two species of
particles organize in two distinct clusters, i.e the particles of the
same species attract each other and the particles of different species
repel each other.

Let us finally comment on the inequivalence of microcanonical and
canonical ensembles for the point vortex system (see also Appendix
\ref{sec_heur}). For ordinary point vortices whose dynamics is governed by
Hamiltonian equations, the statistical equilibrium state is obtained
by maximizing the Boltzmann entropy $S_{B}$ at fixed energy $E$ and
circulation $\Gamma$ (microcanonical ensemble). For Brownian point
vortices whose dynamics is governed by Langevin equations, the
statistical equilibrium state is obtained by maximizing the Boltzmann
free energy $J_{B}=S_{B}-\beta E$ at fixed circulation $\Gamma$
(canonical ensemble). Canonical stability implies microcanonical
stability but the reciprocal is wrong. This means that some states
$(E,\beta)$ that are accessible in the microcanonical ensemble (maxima
of entropy $S$) may be inaccessible in the canonical ensemble (they
are not maxima of free energy $J$). When this happens, we speak of
{\it ensemble inequivalence}. At equilibrium, ensembles are equivalent
for vortices in a disk (or other simple domains) when the Hamiltonian
is postulated to be the only constraint
\cite{es,ca2}. However, they can be inequivalent in other kinds of
domains \cite{ca2} or also in a disk when the conservation of the
angular momentum is taken into account
\cite{son,kl}. Out-of-equilibrium, the kinetic equations describing
Hamiltonian (microcanonical) and Brownian (canonical) point vortices
are {\it very} different in any case. For Hamiltonian vortices, when
$N\rightarrow +\infty$, the kinetic equations reduce to the 2D Euler
equation and finite $N$ effects are necessary to get the convergence
to the microcanonical Boltzmann distribution with temperature
$\beta(E)$
\cite{bbgky}. The relaxation time is larger than $Nt_{D}$ and its
precise scaling with $N$ is not known (we are not even sure that the
system will reach thermal equilibrium).  For Brownian vortices, when
$N\rightarrow +\infty$, the kinetic equations reduce to the mean-field
Fokker-Planck equations (\ref{mf2}) or (\ref{app3})-(\ref{app5}). They
converge towards the canonical Boltzmann distribution with the bath
temperature $\beta$ on a diffusive timescale $L^{2}/\nu$ independent
on $N$. Therefore, although the equilibrium states correspond to mean
field Boltzmann distributions in the two ensembles, the convergence to
these steady states is {\it radically different} in the Hamiltonian
and Brownian descriptions. Thus, our study explicitly shows the
fundamental analogies and differences that exist between a
microcanonical and a canonical description of point vortices. These
analogies and differences are summarized in Table \ref{comp} where we
compare the kinetic equations of Hamiltonian and Brownian vortices (a
completely similar comparison could be made between Hamiltonian and
Brownian systems of material particles with long-range interactions
\cite{hb1,hb2,hb3,hb4,prep}).

\begin{table*}[ht]
%\begin{tabular}{cc}
%\parbox{12cm}
\begin{tabular}
{||c||c|c||c||c||}%
\hline \hline
   Domain of validity   &    Hamiltonian                           &  Brownian    \\
\hline \hline
  Microscopic     & Deterministic Kirchhoff     & Stochastic Langevin\\
 $N$-body  model    & equations    &  equations \\
 for ${\bf r}_{i}(t)$    & (see Eq. (2) of \cite{bbgky})    & (see Eq. (2))\\
\hline \hline
 Exact      & Liouville equation     & $N$-body  \\
kinetic equation    &  (see Eq. (3) of \cite{bbgky})   &  Fokker-Planck equation   \\
for $P_{N}({\bf r}_{1},...,{\bf r}_{N},t)$    &    &  (see Eq. (6))  \\
\hline \hline
 Exact      & Klimontovich equation   & Stochastic kinetic \\
 kinetic equation  for   & (see Eq. (69) of \cite{bbgky})    &  equation \\
 $\omega_{d}({\bf r},t)=\sum_{i=1}^{N}\gamma \delta({\bf r}-{\bf r}_{i}(t))$    &     &  (see Eq. (42)) \\
\hline \hline
  Kinetic equation for    & Euler-Poisson equation     & Fokker-Planck-Poisson  \\
$\omega({\bf r},t)=N\gamma P_{1}({\bf r},t)$   & (see Eq. (35) of \cite{bbgky})   & equation\\
when $N\rightarrow +\infty$  &     & (see Eq. (25))\\
\hline \hline
 Violent relaxation   & Phenomenological equations    &  \\
of the 2D Euler equation.    &  obtained from the MEPP   &  \\
Kinetic equations for  &  (see Eq. (15) of \cite{rsmepp}    &  \\
the coarse-grained   & and Eqs. (23)-(30) of \cite{csmepp})      &  \\
vorticity $\overline{\omega}({\bf r},t)$ &  or kinetic equation  obtained    & \\
 &  from a quasilinear theory    &\\
 &   (see Eq. (18) of \cite{quasi})    &  \\
\hline \hline
Diffusion coefficient for & See Eq. (8) of \cite{rr}, Eq. (4.9)    &  \\
 the violent relaxation  & of \cite{csr} or Eq. (27) of \cite{quasi} & \\
\hline \hline
 Integrodifferential kinetic   & See Eq. (54) of \cite{bbgky}  & unknown \\
equation for $\omega({\bf r},t)$  & (collective effects neglected)     &  \\
at the order $O(1/N)$  &     &  \\
\hline \hline
 Kinetic equation at   & See Eq. (11) of \cite{dn}    & unknown \\
  the order  $O(1/N)$   &  (with collective effects)  and   &  \\
  for axisymmetric flows  &  Eq. (48) of \cite{bbgky}     &  \\
 &   (collective effects neglected)  &  \\
 \hline \hline
Test particle in a   & Fokker-Planck equation  &  \\
(thermal)  bath    &  with a {\it fixed} potential.   &  \\
   &  See Eq. (141) of \cite{bbgky}  &  \\
 & or Eq. (140) of \cite{bbgky} (thermal)   &  \\
\hline \hline
Diffusion coefficient   & See Eq. (102) of \cite{bbgky}   &  \\
for a test particle     &  or Eq. (136) of \cite{bbgky} (thermal)    &  \\
in a (thermal)  bath   &   &  \\
\hline \hline
\end{tabular}
\vspace{.3cm}
%\parbox{5cm}
{\caption{Summary of the different kinetic equations for Hamiltonian
and Brownian vortices. A similar Table would apply to Hamiltonian
and Brownian systems of material particles with long-range
interactions \cite{hb1,hb2,hb3,hb4,prep}.  } \label{comp}}
%\end{tabular}
\end{table*}

\begin{table*}[ht]
%\begin{tabular}{cc}
%\parbox{12cm}
\begin{tabular}
{||c||c|c||c|c||}%
\hline \hline
 & $\beta<\beta_{c}^{-}$ & $\beta<0$  &  $\beta>0$  &  $\beta>\beta_{c}^{+}$ \\
\hline \hline
$2D$ self-gravitating    &     &   & $(m,m)$ attract &  Collapse:  \\
Brownian particles  & & & $\Rightarrow$ cluster &  $\Rightarrow$ Dirac peak  \\
 & & &  &  of mass $Nm$ \\
\hline \hline
$2D$ neutral & & & $(+e,+e)$ repel & Collapse: \\
plasma  & & & $(-e,-e)$ repel & $N/2$ singular \\
        & & & $(+e,-e)$ attract & ``atoms'' $(+e,-e)$ \\
        & & & $\Rightarrow$ homogeneous  &  \\
        & & & state   &  \\
\hline \hline
Single species & Collapse: &  $(\gamma,\gamma)$ attract & $(\gamma,\gamma)$ repel &  \\
$2D$ Brownian  & $\Rightarrow$ Dirac peak of  & $\Rightarrow$ cluster & $\Rightarrow$ concentration   & \\
vortices  &  circulation $N\gamma$ &  (monopole) &  at the boundary & \\
\hline \hline
Multi species & Collapse: &  $(+\gamma,+\gamma)$ attract & $(+\gamma,+\gamma)$ repel & Collapse:  \\
$2D$ Brownian  & singular  &  $(-\gamma,-\gamma)$ attract & $(-\gamma,-\gamma)$ repel & $N/2$ singular   \\
vortices & macroscopic  &  $(+\gamma,-\gamma)$ repel & $(+\gamma,-\gamma)$ attract & microscopic   \\
& dipole  & $\Rightarrow$ two clusters  &  $\Rightarrow$ homogeneous &  dipoles $(+\gamma,-\gamma)$  \\
&$(+\frac{N}{2}\gamma,-\frac{N}{2}\gamma)$  & of opposite   & state  &    \\
&  & sign (dipole) &  &    \\
\hline \hline
\end{tabular}
\vspace{.3cm}
%\parbox{5cm}
{\caption{Schematic ``phase diagram'' showing the analogies between
$2D$ self-gravitating systems (or bacterial populations \cite{crrs}),
$2D$ Coulombian plasmas and $2D$ point vortices in the canonical
ensemble
\cite{diff}. For a system of $N$ self-gravitating Brownian particles with mass
$m$: $\beta_{c}^{+}=4/(GMm)$. For a neutral plasma with $N/2$ charges
$+e$ and $N/2$ charges $-e$: $\beta_{c}^{+}=2/e^2$. For a system of
$N$ Brownian point vortices with circulation $\gamma$:
$\beta_{c}^{-}=-8\pi/(N\gamma^{2})$. For a multi-species system of
Brownian point vortices with $N/2$ vortices $+\gamma$ and $N/2$
vortices $-\gamma$: $\beta_{c}^{-}=-16\pi/(N\gamma^2)$ and
$\beta_{c}^{+}=4\pi/\gamma^2$. In the microcanonical ensemble,
$\beta(E)$ is a function of the energy. There is an equilibrium state
for each value of the energy $E$, so there is no collapse. The
critical temperatures $\beta_{c}^{\pm}$ correspond to $E\rightarrow
\mp\infty$ so the states $\beta<\beta_{c}^{-}$ and
$\beta>\beta_{c}^{+}$ are not accessible in the microcanonical ensemble. }
\label{ana}}
%\end{tabular}
\end{table*}

\appendix

\section{Dissipative guiding center plasma}
\label{sec_gcp}

The equations of motion of a system of charges in a vertical magnetic
field ${\bf B}=B{\bf z}$ are
\begin{eqnarray}
\label{gcp1}
m\frac{d{\bf v}_{i}}{dt}=e_{i}{\bf E}_{i}+e_{i}\frac{{\bf v}_{i}\times {\bf B}}{c},
\end{eqnarray}
where ${\bf E}_{i}=-\nabla_{i}\Phi=\frac{\partial}{\partial {\bf
r}_{i}}\sum_{j\neq i}(2/l)e_{j}\ln x_{ij}$ is the electric field
created self-consistently by the charges \cite{jm}. For large fields
$B\gg 1$, we can neglect the inertial term $d{\bf v}_{i}/dt\simeq {\bf
0}$ and make the guiding center approximation
\begin{eqnarray}
\label{gcp2}
{\bf v}_{i}\simeq -\frac{c}{B^2}{\bf B}\times {\bf E}_{i}.
\end{eqnarray}
The equations of motion are then similar to those of a Hamiltonian 2D
point vortex gas \cite{jm}. For a conservative system (microcanonical
ensemble) both positive and negative temperature states are possible
\cite{onsager}. Let us now consider a {\it dissipative}
magnetized plasma so that the equations of motion can be modeled by
the stochastic process
\begin{eqnarray}
\label{gcp3}
\frac{d{\bf v}_{i}}{dt}=\frac{e_{i}}{m}\left ({\bf E}_{i}+\frac{{\bf v}_{i}\times {\bf B}}{c}\right )-\xi {\bf v}_{i}+\sqrt{2D}{\bf R}_{i}(t),
\end{eqnarray}
where $\xi>0$ is an effective friction coefficient and ${\bf
R}_{i}(t)$ a random force with $\langle {\bf R}_{i}(t)\rangle ={\bf
0}$ and $\langle {R}^{\alpha}_{i}(t){R}^{\beta}_{j}(t')\rangle=\delta_{ij}\delta_{\alpha\beta}\delta(t-t')$. The Einstein relation
reads $\xi=Dm\beta$, implying that the temperature is
necessarily positive for this model. If we consider both a large field limit
$B\rightarrow +\infty$ and a strong friction limit $\xi\rightarrow
+\infty$ in such a way that $\xi\sim eB/(mc)$, we can neglect the
inertial term in Eq. (\ref{gcp3}). Then, after simple manipulations,
we obtain the stochastic equation
\begin{eqnarray}
\label{gcp4}
\left (1+\frac{m^2 c^2}{e_{i}^2 B^2}\xi^2\right ){\bf v}_{i}=\frac{\xi m c^2}{e_{i}B^2}{\bf E}_{i}-\frac{c}{B^{2}}{\bf B}\times {\bf E}+\frac{\xi m^{2}c^2}{e_{i}^{2}B^{2}}\sqrt{2D}{\bf R}_{i}-\frac{m c}{e_{i}B^{2}}\sqrt{2D}{\bf B}\times {\bf R}_{i}.
\end{eqnarray}
Defining ${\bf Q}_i=a_i{\bf R}_i-{\bf B}\times {\bf R}_i$ with
$a_i=\xi m c/e_{i}$, we have $\langle {\bf Q}_{i}(t)\rangle
={\bf 0}$ and $\langle
{Q}^{\alpha}_{i}(t){Q}^{\beta}_{j}(t')\rangle=(a_i^2+1)\delta_{ij}\delta_{\alpha\beta}\delta(t-t')$. Therefore,
the stochastic equation (\ref{gcp4}) can be written in the form of
Eq. (\ref{se1}) provided that the parameters are properly
reinterpreted. If we consider a dissipative plasma, without imposed
magnetic field (${\bf B}={\bf 0}$), the stochastic equation (\ref{gcp4}) obtained in the overdamped limit becomes
\begin{eqnarray}
\label{gcp4bis}
\frac{d{\bf r}_{i}}{dt}=\frac{e_{i}}{m\xi}{\bf E}_{i}+\sqrt{\frac{2D}{\xi^{2}}}{\bf R}_{i}(t).
\end{eqnarray}
Introducing a mobility $\mu=1/(m\xi)$ and a diffusion coefficient
$\nu=D/\xi^{2}$, this yields the stochastic equation (\ref{se1})
without the rotational term. This can provide a microscopic
model of Debye-H\"uckel electrolytes in a canonical setting \cite{dh} which is
valid in $d=2$ or $d=3$ dimensions.

\section{Correlation functions}
\label{sec_id}

Considering the correlation function of the exact vorticity field
(\ref{exact1}), and introducing the one and two-body distributions, we
find that
\begin{eqnarray}
\label{id0}
\langle \omega_d({\bf r})  \omega_{d}({\bf r}')\rangle=\langle \gamma^2 \sum_{i,j}\delta({\bf r}-{\bf r}_{i})\delta({\bf r}'-{\bf r}_{j})\rangle= \langle \gamma^2 \sum_{i=1}^{N}\delta({\bf r}-{\bf r}_{i})\delta({\bf r}'-{\bf r})\rangle\nonumber\\
+\langle \gamma^2 \sum_{i\neq j}\delta({\bf r}-{\bf r}_{i})\delta({\bf r}'-{\bf r}_{j})\rangle = N \gamma^2 P_{1}({\bf r}) \delta({\bf r}-{\bf r}')
+N(N-1) \gamma^2 P_{2}({\bf r},{\bf r}').
\end{eqnarray}
We now assume that the system is spatially homogeneous in average,
with a vorticity distribution $\omega=N\gamma P_0$. Writing $\delta\omega({\bf
r})=\omega_{d}({\bf r})-{\omega}$, we get
\begin{eqnarray}
\label{id1}
\langle \delta\omega({\bf r})\delta\omega({\bf r}')\rangle =\langle \omega_{d}({\bf r})\omega_{d}({\bf r}')\rangle -{\omega}^{2}.
\end{eqnarray}
Starting from  the identity
\begin{eqnarray}
\label{id2}\langle \omega_d({\bf r})  \omega_{d}({\bf r}')\rangle=N\gamma^2 P_{0}\delta({\bf r}-{\bf r}')
+N(N-1)\gamma^2 P_{2}({\bf r},{\bf r}'),
\end{eqnarray}
and introducing the correlation function $h({\bf r}-{\bf r}')$ through the defining relation  $P_{2}({\bf r},{\bf r}')=P_{0}^{2}(1+h({\bf r}-{\bf r}'))$, we obtain for $N\gg 1$:
\begin{eqnarray}
\label{id3}\langle \delta\omega({\bf r})\delta\omega({\bf r}')\rangle =\gamma {\omega}\delta({\bf r}-{\bf r}')+{\omega}^{2}h({\bf r}-{\bf r}').
\end{eqnarray}
The Fourier transform (spectrum) of the vorticity correlations is
\begin{eqnarray}
\label{id4}\langle\delta\hat{\omega}_{\bf k}\delta\hat{\omega}_{\bf k'}\rangle=\frac{\gamma\omega}{(2\pi)^{2}}\left \lbrack 1+(2\pi)^{2}\frac{\omega}{\gamma}\hat{h}(k)\right \rbrack \delta({\bf k}+{\bf k}').
\end{eqnarray}
Finally, using the result (\ref{dh3}), we obtain
\begin{eqnarray}
\label{id5}\langle\delta\hat{\omega}_{\bf k}\delta\hat{\omega}_{\bf k'}\rangle=\frac{n\gamma^2}{(2\pi)^{2}} \frac{1}{1+(2\pi)^{2}\beta n\gamma^2 \hat{u}(k)} \delta({\bf k}+{\bf k}').
\end{eqnarray}

\section{A heuristic relaxation equation for point vortices in the microcanonical ensemble}
\label{sec_heur}

As explained in the Introduction, the dynamical models introduced in
this paper correspond to a canonical description of point vortices. In
these models, the inverse temperature $\beta$ is fixed and the energy
$E(t)$ is not conserved. Following the procedure of \cite{rsmepp,kin,gen}, we
can introduce heuristically a dynamical model of point vortices
corresponding to a microcanonical description by letting the
temperature $\beta(t)$ depend on time so as to rigorously conserve the
energy. Therefore, we now assume in Eq. (\ref{sto2}) that $\mu(t)=\nu\beta(t)$
is a function of time. It is determined by writing $\dot E=\int H
\partial_t P_N d{\bf r}_1...d{\bf r}_N=0$ and using Eq. (\ref{ea1}). This
yields 
\begin{equation}
\beta(t)=-\frac{\int\sum_{i=1}^{N}\frac{\partial P_N}{\partial {\bf r}_i}\cdot \frac{\partial U}{\partial {\bf r}_i}d{\bf r}_1...d{\bf r}_N}{\gamma^2\int\sum_{i=1}^{N}P_N\left (\frac{\partial U}{\partial {\bf r}_i}\right)^2 d{\bf r}_1...d{\bf r}_N}.
\label{heur1}
\end{equation}

In the mean field limit $N\rightarrow +\infty$, we obtain an equation of the form 
\begin{equation}
\frac{\partial\omega}{\partial t}+{\bf u}\cdot\nabla\omega=\nu \nabla\cdot \left (\nabla\omega+\beta(t)\gamma\omega\nabla\psi\right ),
\label{heur2}
\end{equation}
where $\beta(t)$ is determined by the condition $\dot E=\int \psi \partial_t\omega\, d{\bf r}=0$ yielding
\begin{equation}
\beta(t)=-\frac{\int \nabla\omega\cdot \nabla\psi\, d{\bf r}}{\int\gamma\omega (\nabla\psi)^2\, d{\bf r}}.
\label{heur3}
\end{equation}
We can then easily establish (see, e.g., \cite{gen}) that
\begin{equation}
\dot S=\nu\int \frac{1}{\gamma\omega}(\nabla\omega+\beta\gamma\omega\nabla\psi)^2\, d{\bf r}.
\label{heur4}
\end{equation}
The relaxation equation (\ref{heur2}) with the constraint
(\ref{heur3}) has the following properties: (i) $\Gamma$ and $E$ are
conserved (ii) $\dot S\ge 0$ (iii) $\dot S=0$ iff
$\omega=Ae^{-\beta\gamma\psi}$ (iv) a steady state is linearly
dynamically stable iff it is a (local) maximum of $S$ at fixed
circulation and energy. By Lyapunov's direct method, we know that if
$S$ is bounded from above, Eq. (\ref{heur2}) will relax towards a
(local) maximum of $S$ at fixed circulation and energy (if several
local entropy maxima exist, the choice of the entropy maximum will
depend on a notion of basin of attraction). Therefore, this relaxation
equation tends to the microcanonical statistical equilibrium. This
equation was introduced in \cite{kin} by using a maximum entropy
production principle (MEPP) and it is a particular case of the general
class of nonlinear mean field Fokker-Planck (NFP) equations introduced
in
\cite{gen,nfp}. Note that a similar heuristic microcanonical model has
been introduced for self-gravitating systems in the form of  Kramers
and Smoluchowski equations with a time dependent temperature
\cite{csr,sc}.  We stress, however, that Eqs. 
(\ref{heur2})-(\ref{heur3}) do {\it not} describe the real dynamics of
point vortices in the microcanonical ensemble. The exact kinetic
theory of point vortices is the one developed in \cite{bbgky} and it
leads to very different equations. Therefore,
Eqs. (\ref{heur2})-(\ref{heur3}) have probably no physical
justification. However, they can be used as a numerical algorithm to
construct statistical equilibrium states of point vortices in the
microcanonical ensemble (note in this respect that it is possible to
generalize the MEPP so as to include the conservation of angular
momentum in a disk or in an infinite domain).  Similarly,
Eq. (\ref{mf2}) can be used as a numerical algorithm to construct
statistical equilibrium states in the canonical ensemble. These
remarks give them some practical interest. Note also that the
dynamical evolution of the microcanonical equation (\ref{heur2}) with
the constraint (\ref{heur3}) is very different from the dynamical
evolution of the canonical equation (\ref{mf2}) with fixed
$\beta$. Indeed, in the microcanonical ensemble, there exists a
statistical equilibrium state for any value of the energy $E$ so that
Eq. (\ref{heur2}) with Eq. (\ref{heur3}) will relax towards this
equilibrium state. By contrast, in the canonical ensemble, there is no
equilibrium state for $\beta<\beta_c$ and, in that case, the
relaxation equation (\ref{mf2}) leads to vortex collapse.

For the multi-species system, the heuristic microcanonical model in the mean field limit is 
\begin{equation}
\frac{\partial\omega_{\alpha}}{\partial t}+{\bf u}\cdot\nabla\omega_{\alpha}=\nu \nabla\cdot \left (\nabla\omega_{\alpha}+\beta(t)\gamma_{\alpha}\omega_{\alpha}\nabla\psi\right ),
\label{heur5}
\end{equation}
where $\beta(t)$ is determined by the condition $\dot E=\int \psi \partial_t\omega\, d{\bf r}=0$ yielding
\begin{equation}
\beta(t)=-\frac{\int \nabla\omega\cdot \nabla\psi\, d{\bf r}}{\int\omega_2 (\nabla\psi)^2\, d{\bf r}},
\label{heur6}
\end{equation}
where $\omega=\sum_{\alpha}\omega_{\alpha}$ and $\omega_2=\sum_{\alpha}\gamma_{\alpha}\omega_{\alpha}$. We can easily establish that
\begin{equation}
\dot S=\nu\sum_{\alpha}\int \frac{1}{\gamma_{\alpha}\omega_{\alpha}}(\nabla\omega_{\alpha}+\beta\gamma_{\alpha}\omega_{\alpha}\nabla\psi)^2\, d{\bf r}.
\label{heur7}
\end{equation}
The relaxation equations (\ref{heur5}) with the constraint
(\ref{heur6}) have the following properties: (i) $\Gamma_{\alpha}=\int
\omega_{\alpha}\, d{\bf r}$ and $E$ are conserved (ii) $\dot S\ge 0$
(iii) $\dot S=0$ iff
$\omega_{\alpha}=A_{\alpha}e^{-\beta\gamma_{\alpha}\psi}$ (iv) a
steady state is linearly dynamically stable iff it is a (local)
maximum of $S$ at fixed circulations $\Gamma_{\alpha}$ and
energy. Therefore, if $S$ is bounded from above, these
equations will relax towards the microcanonical statistical equilibrium
state of the multi-species point vortex gas. These equations can also
be derived from the maximum entropy production principle \cite{prepa}.

\end{document}